# MEASUREMENTS OF STELLAR MAGNETIC FIELDS WITH THE AUTOCORRELATION OF SPECTRA

E.F. Borra and D. Deschatelets

Département de physique, de génie physique et d'optique. Université Laval

# ABSTRACT

We present a novel technique that uses the autocorrelation of the spectrum of a star to measure the line broadening caused by the modulus of its average surface magnetic field. The advantage of the autocorrelation comes from the fact that it can detect very small spectral line broadening effects because it averages over many spectral lines and therefore gives an average with a very high signal to noise ratio. We validate the technique with the spectra of known magnetic stars and obtain autocorrelation curves that are in full agreement with published magnetic curves obtained with Zeeman splitting. The autocorrelation also gives less noisy curves so that it can be used to obtain very accurate curves. We degrade the resolution of spectra of these magnetic stars to lower spectral resolutions where the Zeeman splitting is undetectable. At these resolutions, the autocorrelation still gives good quality curves, thereby showing that it can be used to measure magnetic fields in spectra where the Zeeman splitting is significantly smaller than the width of the spectral line. This would therefore allow observing magnetic fields in very faint Ap stars with low-resolution spectrographs, thereby greatly increasing the number of known magnetic stars. It also demonstrates that the autocorrelation can measure magnetic fields in rapidly rotating stars as well as weak magnetic fields that give a Zeeman splitting smaller than the intrinsic width of the spectral lines. Finally, it shows that the autocorrelation can be used to find unknown magnetic stars in low resolution spectroscopic surveys.

## 1. INTRODUCTION

The modulus of the average surface magnetic field of a star is usually evaluated by measuring the separations between the Zeeman components of spectral lines (Mathys et al. 1997). This cannot be done for most stars because the Zeeman splitting caused by the average surface magnetic field is usually substantially smaller than the intrinsic width of the lines. A very small fraction of stars have strong magnetic fields where the Zeeman splitting is comparable to the intrinsic width of the lines so that the surface magnetic fields can be measured. However, this requires high-resolution spectrographs and large optical telescopes where it is difficult to have access to observing time. Also, even with large telescopes, high-resolution spectrographs cannot observe faint stars. Furthermore, even with high-resolution spectrographs, the magnetic curves obtained from Zeeman splitting are usually noisy so that it would be useful to have a technique that gives better quality curves. The rotational velocity of stars causes another problem when using Zeeman splitting because the Doppler Effect broadens the lines. Consequently, in practice, surface magnetic fields cannot be detected with Zeeman splitting for stars that have values of the projected rotational velocity *vsini* significantly greater than 10 km/sec. We presently live at the beginning of a new era of large astronomical surveys that obtain low resolution spectra for a very large number of stars. For example, the Sloan Digital Sky Survey (Abazajian et al. 2005) obtains spectra with a resolving power $R = \lambda/\Delta\lambda \sim$ *2000* for hundreds of thousands of stars as faint as *g = 20*. This extremely low resolution does not allow detections of magnetic fields with Zeeman splitting. Other surveys have higher resolutions (e.g. the Gaia-ESO survey, Gilmore et al. 2012) that however are still too small to detect magnetic fields with Zeeman splitting.

The discussion in the previous paragraph shows that it is useful to have techniques that allow the detection of magnetic fields in situations where the Zeeman splitting is smaller than the width of the spectral lines as is the case for weak magnetic fields, rapidly rotating stars or observations with low-resolution spectrographs. We propose the application of the autocorrelation function to stellar spectra to overcome this problem. The advantage of the autocorrelation comes from the fact that it makes an average over all of the spectral lines in the spectral region used. This gives very high signal to noise ratios that allow to measure very small line broadenings. As we shall see, the autocorrelation also easily allows removing a substantial fraction of the effect of photon noise. Another advantage is that the autocorrelation can easily be performed using commercial software (e.g. matlab). Finally, the autocorrelation technique is simpler to perform than other techniques (e.g. Least Square Deconvolution (Donati et al. 1997, Kochukhov, Makaganiuk & Piskunov 2010) and Principal Component Analysis (Martinez Gonzalez et al. 2008)), thereby simplifying searches for magnetic stars in astronomical surveys. Least Square Deconvolution and Principal Component Analysis enhance the signal using the data itself and some approximations. Least Square Deconvolution assumes that all spectral lines have a common shape and they only differ in their scaling factors. Principal Component Analysis assumes that the spectral lines can be described by a linear decomposition in a data base of eigenprofiles and that only a few of them are informative. Once the signal is enhanced, then it uses inversion techniques to infer the magnetic field.

In summary, as we shall see, the autocorrelation allows to measure the shapes of magnetic curves with a very high precision, observe known magnetic stars with low resolution spectra, find stars that have very weak magnetic fields, measure magnetic fields in rapidly rotating stars and find magnetic stars in low spectral resolution astronomical surveys. In this paper, we only introduce and validate the technique. Further work will certainly improve it.

In this work we only consider the application of the autocorrelation to line profiles to obtain the modulus of the average surface magnetic field of the star. The technique should also be applicable to the Stokes V (circular polarization) as well as Q and U (linear polarization) parameters.

## 2. MEASUREMENT OF STELLAR MAGNETIC FIELDS WITH THE AUTOCORRELATION OF THE SPECTRUM

We use the autocorrelation of the intensity $I(\nu)$ as a function of frequency $\nu$ of the spectrum given by

$$I \otimes I = \int_{-\infty}^{\infty} I(\nu+\nu')I(\nu)d\nu \quad , \qquad (1)$$

that makes an average over all the spectral lines in the spectral range used, thereby greatly increasing the signal to noise ratio. The high signal to noise ratio then allows the detection of magnetic fields in spectra where they are totally undetectable with Zeeman splitting. One must however be careful to use spectral regions that do not contain overly broad lines, like the Balmer lines in A stars, otherwise they dominate the autocorrelation and render it useless. In performing the autocorrelation with Equation 1, we subtract the continuum from the spectrum so that the autocorrelation is only performed on the spectral lines. The continuum is removed to subtract its contribution to the autocorrelation curve so that the contribution of the spectral lines becomes easier to detect. The continuum is evaluated with a smoothing function that smoothes the spectrum with a running average that has a spectral width considerably larger than the width of a spectral line. The autocorrelation symmetrizes the spectral lines but this is not a problem for the measurements of magnetic fields since the Zeeman effect introduces a symmetric broadening. This is also not a problem for Doppler broadening caused by the rotation of the stars since it also causes a symmetrical broadening. An advantage of the autocorrelation technique is that the autocorrelation can easily be performed with commercial software. For example, we perform it with the Matlab function *xcorr*.

Besides broadening from the magnetic fields, there is also the intrinsic width of the line, broadening from the instrumental profile, broadening caused by turbulence, thermal broadening from the temperature of the gas and broadening due to *vsini*, the projected rotational velocity of the star. If we exclude strong lines, such as Balmer or Helium lines, the intrinsic width of the lines has a small effect. Instrumental broadening can be removed with deconvolution techniques. In practice, turbulence and thermal broadenings are minor effects but rotation can cause a very large broadening that totally dominates over Zeeman splitting. Values of *vsini* of 10 km/sec are already significant. It is therefore very important to take rotational broadening into account. The fact that the broadenings due to rotation and magnetic fields do not have the same wavelength and frequency dependence allows us to separate them. We prefer to work in frequency units because, in a spectrum defined in frequency units, the broadening due to magnetic fields is independent of frequency, while thermal, turbulence and rotational broadenings increase linearly with *vsini* and frequency. To remove the linear effects (e.g. *vsini* ) from the magnetic field effect, we compute the autocorrelation at the 5 different frequency ranges in Table I. The 5 spectral ranges were chosen so that they run over a large frequency range, contain a large number of lines and also exclude the Hydrogen Balmer lines which would otherwise have dominated the autocorrelation because of their large widths. We also avoided the regions at the lowest and highest frequencies because they are significantly noisier. They were also chosen to exclude regions where there were instrumental problems in the spectral data that we retrieved from publicly available databases. On the other hand, it is not necessary to remove the effect of rotational velocities for stars that have low rotational velocities and strong magnetic fields, like most known magnetic Ap stars. It is then preferable to use a single spectral range that includes many lines and avoids strong lines. The figure at the left of Figure 1 shows such a spectral region of the known magnetic star HD 144897 and the figure at the right of Figure 1 shows the central region of the autocorrelation of this spectral region. The fact that the profile of the autocorrelation is very smooth clearly shows that the signal to noise ratio is greatly increased by the autocorrelation. We choose the width at 25% peak intensity, identified by the rectangle at the bottom of the figure at the right in Figure 1, to quantify the width of the autocorrelation. There probably are better ways that use the entire profile to quantify the width of the autocorrelation (e.g. least-square fitting the entire profile) but in this work we only consider this simple technique. We then added, with software, a considerable quantity of Gaussian noise to the spectrum in Figure 1 to generate the spectrum in Figure 2. The autocorrelation of the very noisy spectrum in Figure 2 illustrates another advantage of the autocorrelation because we can see that it separates noise from the rest of the signal to generate the sharp peak in the plot at the right. If we remove the sharp peak and evaluate the width of the autocorrelation at a 25% peak intensity that starts at the new peak (0.52) we find the same width as in Figure 1. This demonstrates that we can remove a considerable fraction of the effect of photon shot noise from the autocorrelation.

Figure 3 shows a plot of the width of the autocorrelation as a function of frequency for a computer simulation of a star that has no magnetic field but a rotational velocity of 15 km/sec. The computer simulated spectrum is created by generating 1000 spectral lines, in the spectral range 3526 < $\lambda$ < 9214 Angstroms that have a Gaussian shape with a dispersion that generates a line having a full width at half maximum equal to the width predicted by *vsini* = 15 km/sec. The lines are placed at random locations within the spectral range. The peak intensities of the lines also vary at random between 0.0 and 1.0. The spectral range used is the spectral range of the FEROS ESO spectrograph, which is the spectrograph used to analyze many of the stars in the figures displayed in sections 4 and 5. Figure 3 shows that the width of the autocorrelation increases linearly with frequency. Figure 4 shows the width of the autocorrelation as a function of frequency for a computer simulation for the same star that has a rotational velocity of 15 km/sec and also a surface magnetic field of 15 kGauss. The computer simulated spectrum was created with the same technique used to generate Figure 3. We model the Zeeman splitting by assuming that all the spectral lines are split in triplets with a central π component and σ+ and σ- components. The Landé g-factor of the lines is set to *g=1.4* for all the lines, in agreement with the average Landé g-factor of the lines in known magnetic stars. Using a range of values of the g-factor will have a negligible effect because the autocorrelation gives an average of the widths of the lines, which are proportional to the g-factors and, furthermore, the range of g-factors is small for the majority of the lines. Figure 4 shows the effect of the magnetic field that introduces a broadening that is independent of frequency. This changes the slope of the linear fit in Figure 3, where there is no magnetic field widening, and therefore also the mean value of the autocorrelation of the five regions. To intuitively understand this, consider the simple modeling of the line profiles by a Gaussian function. If we start from the case where the rotational velocity is large enough that the broadening due to *vsini* dominates, we have a Gaussian with a standard deviation $\sigma$ that varies linearly with frequency; consequently the full width at half maximum, given by $2.335\sigma$, varies linearly with frequency. If we add a magnetic field, it adds a constant proportional to the strength of the magnetic field and independent of *vsini* to the standard deviation $\sigma$, the effect of which is to decrease the slope of the linear relationship. This can readily be understood by first considering the case where the magnetic field is negligible and *vsini* large; the standard deviation then varies linearly with frequency and a steep slope. At the other extreme, the magnetic field is very strong and dominates over *vsini*; the standard deviation is then nearly independent of frequency. Consequently a gradual increase of the magnetic field will lead from the first extreme to the second one. In section 3 we use the mean value of the widths of the autocorrelations of the five regions to quantify the effect of the magnetic field.

In the next three sections, we validate the use of the autocorrelation technique and its application to the study of magnetic fields in stars that are known to be magnetic, to measurements of magnetic fields in rapidly rotating stars and to the search for magnetic stars in spectroscopic surveys.

## 3. HIGH PRECISION MEASUREMENTS OF MAGNETIC FIELDS IN KNOWN MAGNETIC STARS

In this section we measure the magnetic fields of known magnetic stars with the autocorrelation technique to first validate it and then to demonstrate that it can be used to measure with a very high precision the shapes of the time variations of the modulus of the average surface magnetic fields of known magnetic stars, thereby allowing us to study them in great details.

The moduli of the surface magnetic fields are usually obtained by measuring the Zeeman splitting (Mathys et al. 1997). The Zeeman splitting cannot therefore be significantly smaller than the intrinsic width of the lines or the widening caused by *vsini*. In practice this limits one to measurements of strong surface magnetic fields; consequently one can measure the variations of the modulus of the surface magnetic fields in only a small number of magnetic stars that have strong magnetic fields. Furthermore, there are significant uncertainties coming from photon shot noise and blending effects occurring from nearby spectral lines.

To validate the use of the autocorrelation, we used the magnetic stars in Mathys et al. (1997) that gives magnetic curves for several known magnetic stars. We found several spectra of these stars on the European Southern Observatory website. The spectra were obtained with the HARPS spectrograph, have a spectral range extending between 3780 and 6900 Angstroms and have a high spectral resolving power of $R = \lambda/\Delta\lambda = 115,000$. In this section the autocorrelation is carried out in a single spectral range because we use it to measure magnetic fields in magnetic stars that have very small rotational velocities, thereby rendering the effect of *vsini* negligible. This is discussed in section 2 and the spectral range used can be seen in Figure 1. Figures 5 to 7 compare the autocorrelation curves, obtained from spectra having the original spectral resolving power R = 115,000, to the modulus of the surface magnetic field curves for three stars observed by Mathys et al. (1997). The figures at the left show the width of the autocorrelation as a function of phase while the figures at the right show the surface magnetic field as a function of phase from Mathys et al. (1997). We can see that the autocorrelation gives curves having shapes and amplitudes similar to the magnetic curves. However, the autocorrelation gave less noisy curves, thereby illustrating one of the advantages of the autocorrelation which averages over many lines and therefore gives a profile that has a very high signal to noise ratio. An estimate of the random noise in our curves is given by the error bars (one standard deviation length) in Figures 5 to 7 that give an estimate of the standard deviation obtained by least square fitting a cosine to the autocorrelation curve. The dotted lines in the figures show the cosine curves.

Mathys et al. (1997) use the Fe II line at 6149 Angstroms to measure the modulus of the magnetic field. This is a spectral line which has a very large Landé *g*-factor (g=2.7) and, furthermore, only has 2 Zeeman components, while the majority of the spectral lines have several Zeeman components. It is therefore the spectral line which is the most sensitive to the Zeeman effect. Figure 2 in Mathys et al. (1997) shows the Zeeman splitting of that Fe II line for the magnetic stars that they observed. We can see that, although the Fe II line of the stars in Figures 5 to 7 (HD 81009, HD 93507 and HD 144897) have a very large Zeeman splitting, the magnetic field curves obtained by Mathys et al. (1997) are considerably noisier than the curves given by the autocorrelation. Figures 5 to 7 therefore demonstrate that the autocorrelation can reproduce known magnetic field curves with a higher precision than Zeeman splitting. This is particularly apparent for HD 93507 in the region surrounding the phase = 0.5 in Figure 6. Mathys et al. (1997) write that this is caused by blending with spectral lines from other chemical elements that vary in strength with phase, and are stronger at this particular phase. We obtained even less noisy autocorrelation curves by using the Wiener deconvolution technique discussed in section 4.

## 4. MEASUREMENTS OF MAGNETIC FIELDS IN LOW-RESOLUTION SPECTRA, RAPIDLY ROTATING STARS AND MEASUREMENTS OF WEAK MAGNETIC FIELDS.

The major advantage of the autocorrelation is that it allows measuring the moduli of magnetic fields that are not measurable with Zeeman splitting because the splitting is smaller than the width of the spectral lines. The autocorrelation therefore allows measuring the modulus of weak magnetic fields. It also allows measurements of the magnetic fields of known magnetic stars with low resolution spectra, while Zeeman splitting necessitates high-resolution spectra. Also, the rotational velocity of the star broadens the spectral lines so that the conventional techniques do not allow to measure the modulus of magnetic fields of magnetic stars that have values of *vsini* greater than a few km/sec. The autocorrelation technique allows to measure surface magnetic fields in stars that have much larger values of *vsini*.

To demonstrate these advantages, we degrade with software the R = 115,000 resolving power of the spectra used in section 3 and use the autocorrelation to obtain magnetic curves in the degraded spectra. The spectral resolution is degraded by first convolving the spectrum with a Gaussian function having a full width at half-maximum equal to the desired spectral resolution. For instance, if we want to degrade it by a factor of 2, we convolve the original spectrum with a Gaussian having a full width at half-maximum equal to twice the value of the original spectral resolution. We then decrease the number of spectral samples to the appropriate number. For example, in the case where we reduce the spectral resolution by a factor of 2, we start from the original spectrum that has a total of 189,000 samples, convolve it with the Gaussian, and interpolate the intensities to generate a new spectrum having 94,500 samples. Figure 8 gives, for HD 81009 (the star in Figure 5), the autocorrelation curves obtained from spectra degraded by factors of 5 (R= 23,000), 10 (R= 11,500) and 25 (R= 4,600). The degradation increases left to right in Figure 8. The bottom figures give the degraded profiles of the Fe II line at 6149 Angstroms which was used to measure the magnetic fields in Figures 5, 6 and 7 by Mathys et al. (1997). We can see that the autocorrelations from the first two degraded spectra (R= 23,000 and R= 11,500) clearly show the magnetic variation and give curves similar to the magnetic curve of Mathys et al. (1997) in Figure 5, but the last one (R= 4,600) does not. In the figures at the bottom, the Zeeman splitting is barely visible in the Fe II line at 6149 Angstroms degraded by a factor of 5 (R= 23,000) and is totally undetectable for factors of 10 (R= 11,500) and 25 (R= 4,600). We then deconvolved the instrumental profiles of the degraded spectra using a Wiener filter and applied again the autocorrelation. Figure 9 shows the autocorrelation curve for the spectra degraded by a factor of 25 (R =4,600), which is the worst case in Figure 8, after we deconvolved the instrumental profiles with the Wiener filter. We can see a considerable improvement since the autocorrelation curve is similar to the autocorrelation curve in Figure 5. The Zeeman splitting remains undetectable in the profile of the deconvolved Fe II line at 6149 Angstroms at the right of Figure 9 because the deconvolution does not remove the effect of the low sampling.

Our discussion of low resolution spectra demonstrates an interesting application of the autocorrelation because it allows the measurements of the moduli of magnetic fields in Ap stars that are too faint to be observed with the high-resolution spectrographs needed for measurements of the Zeeman splitting. This can be seen in Mathys et al. (1997) who use spectrographs with R > 70,000 and only list 42 stars, 70% of which have V magnitudes brighter than V= 8 and the faintest one has V= 10. Considering the low resolutions used to generate Figure 8 and, in particular Figure 9 (R = 4,600) which used Wiener deconvolution, we can see that the autocorrelation would allow the measurements of magnetic fields in an extremely large number of Ap stars having very faint magnitudes.

We then added Gaussian noise with software to the spectra of the magnetic stars used in Figures 5 to 7 and found that the added noise did not change significantly the results. This is due to the averaging effect of the autocorrelation and the fact that, as explained in section 3, the autocorrelation removes photon noise to generate the sharp peak that can be seen in Figure 2. This is another major advantage of the technique.

The previous sections only discuss spectral profiles that are broadened by the instrumental profile. However the main conclusion, that the autocorrelation detects magnetic fields in broadened spectral lines where Zeeman splitting is undetectable, obviously also applies to any spectral broadening effect. High rotational velocities will also broaden the line profile and make the Zeeman splitting undetectable with conventional techniques. The previous sections therefore demonstrate that the autocorrelation can also be used to detect magnetic fields in stars that have a value of *vsini* that broadens too much the spectral lines to allow the Zeeman splitting to be measured with conventional techniques.

Another major advantage of the autocorrelation, and perhaps the most important one, is demonstrated by the discussion of the degraded spectra; it can detect surface magnetic fields that are too weak to be measured with Zeeman splitting because the splitting is significantly smaller than the intrinsic width of the lines. We did not carry out a detailed demonstration of this advantage because it can readily be seen from our discussion of instrumental broadening, which shows that magnetic fields can be detected even if the Zeeman splitting is considerably smaller than the spectral resolution. This is clearly proven when referring to Figure 8, where the autocorrelation in the middle (R= 11,500) gave a curve comparable to the magnetic curve observed in Figure 5, although the Zeeman splitting is totally undetectable in the broadened profile used. This is even more evident if we compare Figure 9, obtained from spectra degraded by a factor of 25 (R= 4,600) and then Wiener deconvolved, to the magnetic curve in Figure 5.

## 5. FINDING MAGNETIC STARS IN ASTRONOMICAL SURVEYS

We currently live in an epoch where spectroscopic astronomical surveys are becoming increasingly common. The autocorrelation can also be used to find magnetic stars in astronomical surveys. These surveys are usually carried out with low-resolution spectra where, as discussed in section 4, the autocorrelation has major advantages. The problem with low-resolution spectra comes from the fact that the low resolution makes it impossible to measure the Zeeman splitting. On the other hand, as discussed in section 4, the autocorrelation allows us to detect magnetic fields in low resolution spectra where the Zeeman splitting is undetectable.

The autocorrelation measures the average width of the lines; however the average width depends not only from magnetic fields but also from other effects, such as rotational velocities, thermal broadening, broadening due to turbulence and instrumental broadening. Some of the effect of instrumental broadening can be removed using deconvolution techniques such as Wiener deconvolution. However, as can be seen in Figure 9, removing the instrumental broadening does not remove the disadvantages of low resolution spectroscopy that come from its small sampling. The broadening caused by magnetic fields can be distinguished from rotational, thermal, and turbulence broadening because, as discussed in section 2, it is independent of frequency while the two other broadening effects are frequency dependent. As discussed in section 2, to obtain the frequency dependence of the autocorrelation, we perform the autocorrelation in the 5 different intervals of frequencies in table I. Plots of the width of the autocorrelation as function of frequency for stars that do not have a magnetic field (Figure 3) show a linear dependence with a slope that depends on the value of *vsini*. Plots of the width of the autocorrelation as function of frequency for stars that have a magnetic field (Figure 4) show a linear dependence with a slope that depends on both the value of *vsini* and of the strength of the magnetic field because the magnetic field introduces a broadening that is independent of frequency.

We cannot use the width of the autocorrelation alone to find magnetic stars in a survey because the width also depends on broadening from rotational velocities, thermal and turbulence effects as well as the instrumental profile. The instrumental profile will have the same effect on all spectra taken with the same spectrograph and will still allow separating magnetic stars from non-magnetic stars. A lower spectral resolution will only make the separation more difficult. This is discussed in the next paragraphs. In practice, the effect of *vsini* is more important. As discussed in the previous paragraph, for every individual spectrum, the autocorrelation width varies linearly with frequency with a slope that depends on *vsini*. Consequently, we expect that non-magnetic stars, where *vsini* has the major broadening effect, will follow a tight relation in a plot of their slopes of the width of the autocorrelation as a function of frequency versus their mean values of the width. Because adding a magnetic field increases the widths and decreases the slope of the width of the autocorrelation as function of frequency, it places the magnetic stars in locations that are not on the tight relation where the non-magnetic stars are. Consequently, we can use plots of slope versus mean width to find magnetic stars in surveys. Magnetic stars will be identified as stars that are at locations that are at statistically significant positions away from the tight relation where the non-magnetic stars are.

We shall use the mean value of the widths of the five autocorrelations in the five spectral regions in Table I to quantify the effect of the magnetic field. We shall use plots of the slopes of the linear relation in Figure 4 versus the mean widths of the 5 autocorrelations to distinguish magnetic stars from non-magnetic stars. Figure 10 shows a plot of the slopes versus mean widths of the autocorrelation for A stars in the spectral type range between B9.5 and A8 and F stars in the spectral type range between F0 and F8 without magnetic fields. They have values of *vsini* ranging between 8 and 26 km/seconds. The spectra used to generate this figure, as well as the spectra of magnetic stars used to generate the figures that follow, were retrieved from the European Southern Observatory (ESO) database. They were obtained with the FEROS ESO spectrograph in the spectral range $3526 < \lambda < 9214$ Angstroms, with a resolving power of R = 48,000 and with the ELODIE ESO spectrograph in the spectral range $4000 < \lambda < 6800$ Angstroms, with a resolving power of R = 42,000. The 5 regions in Table I are within the FEROS and ELODIE spectral ranges. In Figure 10, we identify the A stars with dots and the F stars with crosses. Note that the A and F stars are within the same range of values of *vsini*. Figure 10 shows linear relations that have different slopes for the two spectral types. It clearly shows that one must use the slopes and widths of a group of stars within the same spectral type range when searching for magnetic stars in spectroscopic surveys. The reason why different spectral type ranges have different slopes comes from the fact that the spectral lines have different positions and intensities within the 5 spectral ranges in Table I. Figure 11 shows a plot of the slopes versus mean widths of the autocorrelation for non-magnetic A stars in the spectral type range B9 to A8 (dots) and magnetic stars (crosses). The dashed line gives a boundary 3 standard deviations away from the linear least squares fit for the non-magnetic stars in the spectral type range B9 to A8. The spectra of the magnetic stars were obtained with the FEROS spectrograph. We can see that all of the magnetic stars can readily be identified. The magnetic fields of the stars in this figure range between 2.4 kG and 15 kG. Consequently magnetic stars having magnetic fields smaller than 2.4 kG could be found. Note also that the resolving powers R = 48,000 and R = 42,000 of the spectra that we used is smaller than the resolving powers (typically R of the order of 100,000) used to measure magnetic fields with Zeeman splitting. If we had used higher resolution spectrographs, the 3 standard deviations limits would have been smaller and allowed detection of magnetic fields considerably smaller than 2 kG. Furthermore, as discussed below, significantly smaller limits could have been reached after deconvolution of the instrumental profile. We did not remove the instrumental profile because we do not know it. We use the distance from the linear relation given by non-magnetic stars to separate magnetic from magnetic stars. The question that then arises is how to find the non-magnetic stars in a survey. This can easily be done by simply assuming that the majority of the stars in the survey have very weak magnetic fields. We know that this assumption is valid because only a very small fraction of the known stars (mostly Ap stars) have a magnetic field that is measurable by Zeeman splitting or circular polarization in high resolution spectra. Consequently, in a survey, one can simply use a random sample of the spectra and assume that they are non-magnetic stars. In our case, to generate the linear relations in Figures 10 to 14, we simply used the spectra of stars that were not known to be magnetic in the European Southern Observatory database.

We then degrade the resolving power by a factor of 5, with the technique described in section 2, to R = 9,600 and perform the autocorrelation on the degraded spectra. Figure 12 shows a plot of the slopes versus mean widths of the autocorrelation for non-magnetic A stars (dots) and magnetic stars (crosses). All of the magnetic stars are beyond the 3 sigma boundary. It is important to note that this is a very low resolution (0.5 Angstroms at a wavelength of 5000 Angstroms), so that the Zeeman splitting of the stars in Figure 12 would be undetectable, as can be seen in the middle figure of Figure 8, where R =11,500. As can be seen in Figure 5, this star (HD 81009) has a strong magnetic field that varies between 7.5 and 9.5 kGauss. Finally, we degrade the spectral resolution by a factor of 10, with the technique described in section 2, to R = 4,800 and perform the autocorrelation on the degraded spectra. Figure 13 shows a plot of the slopes versus mean widths of the autocorrelation for non-magnetic A stars (dots) and magnetic stars (crosses). Only 3 of the magnetic stars are beyond the 3 sigma boundary. For obvious reasons the technique is less capable to detect magnetic stars at this very low spectral resolution, which is only a factor of 2.4 higher than the resolving power of the Sloan Digital Sky Survey. It would therefore still be possible to find rare stars having extremely strong magnetic fields in the SDSS survey.

We then apply the Wiener deconvolution and remove the spectral instrumental profile with the Wiener deconvolution. The Wiener deconvolution can easily be done with the Matlab function *deconvwnr*. Figure 14 shows the results obtained from the autocorrelation of spectra with the same R = 9,600 resolving power that was used to generate Figure 12 but after the Wiener deconvolution was applied to the spectra. We can see a considerable improvement. Figure 15 shows the results obtained from the autocorrelation of spectra with the same R = 4,800 resolving power that was used to generate Figure 13 but after the Wiener deconvolution was applied to the spectra. Once again, the Wiener deconvolution yielded considerably better results. On the other hand, the Zeeman splitting is totally undetectable in the deconvolved Fe II line in Figure 9 (R=4600), because the deconvolution does not remove the effect of the small sampling.

The previous discussions consider detections of magnetic stars with a single spectrum. On the other hand, it is easier to find magnetic fields in stars simply by obtaining the variation of autocorrelation width with time with a few spectra. This can be seen by noting the variations of autocorrelation width in the figures of HD 81009 (Figures 8 and 9). This can particularly be seen in Figure 9 where one can clearly see the variations of the autocorrelation in a spectrum with a resolving power R= 4,800 after Wiener deconvolution. However, Doppler broadening can also be induced by turbulence, so that time varying turbulence could also cause time varying line broadening. Our discussion of Doppler broadening caused by *vsini* shows that it induces a slope that increases with increasing *vsini*, while magnetic broadening induces, in plots of the width of the autocorrelation as a function of frequency, a linear relation with a slope that decreases with increasing magnetic field (see Figures 4 and 5). Consequently magnetic broadening can be distinguished from Doppler broadening because the slope varies proportionally to broadening in the case of Doppler broadening while the contrary occurs for magnetic broadening.

In all the figures, we always plot the width of the autocorrelation. We never convert the autocorrelation width to Gauss units. The width of the autocorrelation depends on several other factors (intrinsic width of the line, broadening from the instrumental profile, thermal broadening, turbulence and broadening due to *vsini* ) that are discussed in detail in this section and the preceding sections. These effects are minor in Figures 5 to 9 because the magnetic fields of these stars are very strong and the values of *vsini* are very small. The effect of these factors depends on the particular application and they can be removed by deconvolution or computer modelling. For example, in the case of observation with low resolution spectra, the main factor is the instrumental profile that can be easily removed with Wiener deconvolution. In the case of measurements of weak magnetic fields in slowly rotating stars, one must also take into account the intrinsic width of the line and thermal broadening. For rapidly rotating stars, the effect of *vsini* will totally dominate. Magnetic broadening can be distinguished from rotational, thermal, and turbulence broadening because, as discussed in section 2, it is independent of frequency while the other broadening effects are frequency dependent. In the case of obtaining accurate magnetic curves in known magnetic stars, one could also use a magnetic field obtained from Zeeman splitting with a single spectrum to calibrate the autocorrelation amplitudes. A discussion of the conversion of the autocorrelation width to Gauss units is complex and beyond the scope of the present article, which simply introduces and validates the technique. This will be discussed in another article.

6. CONCLUSION

Our detailed discussions of the use of the autocorrelation of the spectrum of a star to measure the modulus of its surface magnetic fields show that it has major advantages over conventional techniques that measure the splitting between Zeeman components.

These advantages come from the fact that the autocorrelation makes an average over many spectral lines and therefore gives a measure of the average width of the spectral lines that has a very high signal to noise ratio. This allows the measurements of very small broadening effects. Besides Zeeman broadening, the widths of the spectral lines also depend on other effects. In practice, the instrumental profile and the projected rotational velocity *vsini* are the most important effects to consider. We can separate the effects of rotation, thermal broadening and turbulence from the effect of magnetic field broadening because they do not have the same wavelength and frequency dependences. We work in frequency units (Hertz), rather than wavelength units (Angstroms), because the effects of *vsini*, thermal broadening and turbulence on the autocorrelation vary linearly with frequency while the effect of the magnetic field is independent of frequency.

We begin by validating the use of the autocorrelation by analyzing the spectra of three known magnetic stars. Figures 5 to 7 compare the magnetic curves of these magnetic stars measured with the autocorrelation to curves of magnetic stars measured with Zeeman splitting. This comparison shows that the autocorrelation not only reproduces very well the shapes of the magnetic curve as measured by using Zeeman splitting, but also gives curves that are less noisy. This demonstrates that one can use the autocorrelation to measure shapes of magnetic curves that are more accurate than those measured with Zeeman splitting. Consequently, the autocorrelation can give better estimates of the geometry of the surface of the magnetic field. It can also find (or set limits to) variations of the surface magnetic field that are not periodic. The autocorrelation carries out an average over many spectral lines that translate into a single average spectral line. This gives another advantage of the technique since the distribution of chemical abundances varies over the surface of magnetic Ap stars. Consequently, averaging over many spectral lines gives a better estimate of their surface magnetic fields.

We then consider applications of the autocorrelation to measurements of magnetic fields in spectra where they cannot be measured from Zeeman splitting. To demonstrate these applications, we degraded, with software, the spectral resolution of the spectra of known magnetic stars. The autocorrelation could measure magnetic broadening in significantly degraded spectra. It is important to note that the most important conclusion to draw is not that the autocorrelation can reproduce known magnetic field curves in degraded spectra. The most important conclusion is that the magnetic field variations are obtained from spectra where the magnetic field is unobservable with Zeeman splitting. This validates other interesting uses of the autocorrelation in spectra where the Zeeman splitting is significantly smaller than the width of the spectral lines. It therefore allows the study of the magnetic curves of known magnetic stars with low resolution spectra, the measurement of magnetic fields in rapidly rotating stars that have large values of *vsini* and finding magnetic stars in astronomical surveys. The use of low resolution spectra also gives an interesting application of the autocorrelation to the study of Ap stars because it can measure magnetic fields in very faint stars that cannot be observed with the high-resolution spectrographs needed for measurements of the Zeeman splitting. This would therefore allow the observation of magnetic fields in an extremely large number of Ap stars and thereby allow us to better understand them. Finally, the detection of magnetic fields in degraded spectra, where the Zeeman splitting is smaller than the widths of the spectral lines, also demonstrates what perhaps the most interesting use of the autocorrelation is: Measurements of weak magnetic fields that cannot be measured with Zeeman splitting because the splitting is considerably smaller than the width of the spectral lines.

We therefore conclude that the autocorrelation can measure the modulus of the magnetic field of known magnetic stars with a significantly higher precision and can also measure them with spectra that have a spectral resolution too low for Zeeman splitting measurements. It can also measure magnetic fields that are too weak to be measured with Zeeman splitting so that it can measure them in known magnetic stars that have weak surface fields and can also find new types of magnetic stars. It can measure magnetic fields in rapidly rotating stars. It can find magnetic stars in astronomical surveys that obtain spectra with a low spectral resolution.

ACKNOWLEDGEMENTS

This research has been supported by the Natural Sciences and Engineering Research Council of Canada.

FIGURE CAPTIONS

Figure 1

The figure at the left shows the spectral region used to carry out the autocorrelation for slowly rotating magnetic stars. The figure at the right shows the autocorrelation function of this spectral region for the magnetic star HD 144897.

Figure 2

The figure at the left shows the same spectral region shown in Figure 2 to which we added, with software, a considerable quantity of Gaussian noise. The figure at the right shows the autocorrelation function of this spectral region.

Figure 3

It shows the width of the autocorrelation as a function of frequency for a computer simulation of the spectrum of a star that has no magnetic field but a rotational velocity of 15 km/sec.

Figure 4

It shows the width of the autocorrelation as a function of frequency for a computer simulation of the spectrum of the same star shown in Figure 4 to which we added a surface magnetic field of 15 kGauss.

Figure 5

The figure at the left gives the width of the autocorrelation as a function of phase obtained from spectra having the resolving power R = 115,000, for HD 81009. The dotted line shows a cosine curve least square fitted to the autocorrelation data. The error bar (one standard deviation length) gives an estimate of the standard deviation obtained from the cosine curve. The figure at the right gives the surface magnetic field as a function of phase from Mathys et al. (1997) for the same star.

Figure 6

The figure at the left gives the width of the autocorrelation as a function of phase obtained from spectra having the resolving power R = 115,000, for HD 93507. The dotted line shows a cosine curve least square fitted to the autocorrelation data. The error bar (one standard deviation length) gives an estimate of the standard deviation obtained from the cosine curve. The figure at the right gives the surface magnetic field as a function of phase from Mathys et al. (1997) for the same star.

Figure 7

The figure at the left gives the width of the autocorrelation as a function of phase obtained from spectra having the resolving power R = 115,000, for HD 144897. The dotted line shows a cosine curve least square fitted to the autocorrelation data. The error bar (one standard deviation length) gives an estimate of the standard deviation obtained from the cosine curve. The figure at the right gives the surface magnetic field as a function of phase from Mathys et al. (1997) for the same star.

Figure 8

It gives, for HD 81009 (the star in Figure 5) , the autocorrelation curves for three spectra degraded by factors of 5 (R= 23,000), 10 (R= 11,500) and 25 (R= 4,600). The degradation increases left to right. The bottom figures show the degraded profiles of the Fe II line at 6149 Angstroms which was used to measure the magnetic fields in Figure 6 by Mathys et al. (1997).

Figure 9

It shows the autocorrelation curve obtained from the spectra degraded by a factor of 25 (R =4600) after we deconvolved the instrumental profiles with the Wiener filter. The figure at the right shows the deconvolved Fe II line at 6149 Angstroms.

Figure 10

It shows a plot of the slopes versus mean widths of the autocorrelations for A stars (dots) and F stars (crosses) without magnetic fields. They have values of *vsini* ranging between 8 and 26 km/seconds.

Figure 11

It shows a plot of the slopes versus mean widths of the autocorrelations for non-magnetic A stars (dots) and magnetic stars (crosses). The dashed line gives a boundary 3 standard deviations away from the linear least squares fit for the non-magnetic stars.

Figure 12

It shows a plot of the slopes versus mean widths of the autocorrelations, obtained from spectra degraded to a resolving power $R = 9,600$, for non-magnetic A stars and magnetic stars (crosses). The dashed line gives a boundary 3 standard deviations away from the linear least squares fit for the non-magnetic stars.

Figure 13

It shows a plot of the slopes versus mean widths of the autocorrelations, obtained from spectra degraded to a resolving power $R = 4,800$, for non-magnetic A stars (dots) and magnetic stars (crosses). The dashed line gives a boundary 3 standard deviations away from the linear least squares fit for the non-magnetic stars.

Figure 14

It shows the results obtained from the autocorrelations of spectra with the same $R = 9,600$ resolving power used in Figure 13 but after the Wiener deconvolution was applied to the spectra.

Figure 15

It shows the results obtained from the autocorrelations of spectra with the same $R = 4,800$ resolving power used in Figure 14 but after the Wiener deconvolution was applied to the spectra.

TABLE 1

|        | Frequency range (Hz) | | Wavelength range (Å) | |
|--------|----------------------|----------------------|------|------|
| Region | From | To | From | To |
| 1. | $5.31 \times 10^{14}$ | $5.58 \times 10^{14}$ | 5374 | 5655 |
| 2. | $5.75 \times 10^{14}$ | $6.03 \times 10^{14}$ | 4979 | 5219 |
| 3. | $6.41 \times 10^{14}$ | $6.69 \times 10^{14}$ | 4484 | 4678 |
| 4. | $7.00 \times 10^{14}$ | $7.27 \times 10^{14}$ | 4125 | 4289 |
| 5. | $7.37 \times 10^{14}$ | $7.51 \times 10^{14}$ | 3996 | 4071 |

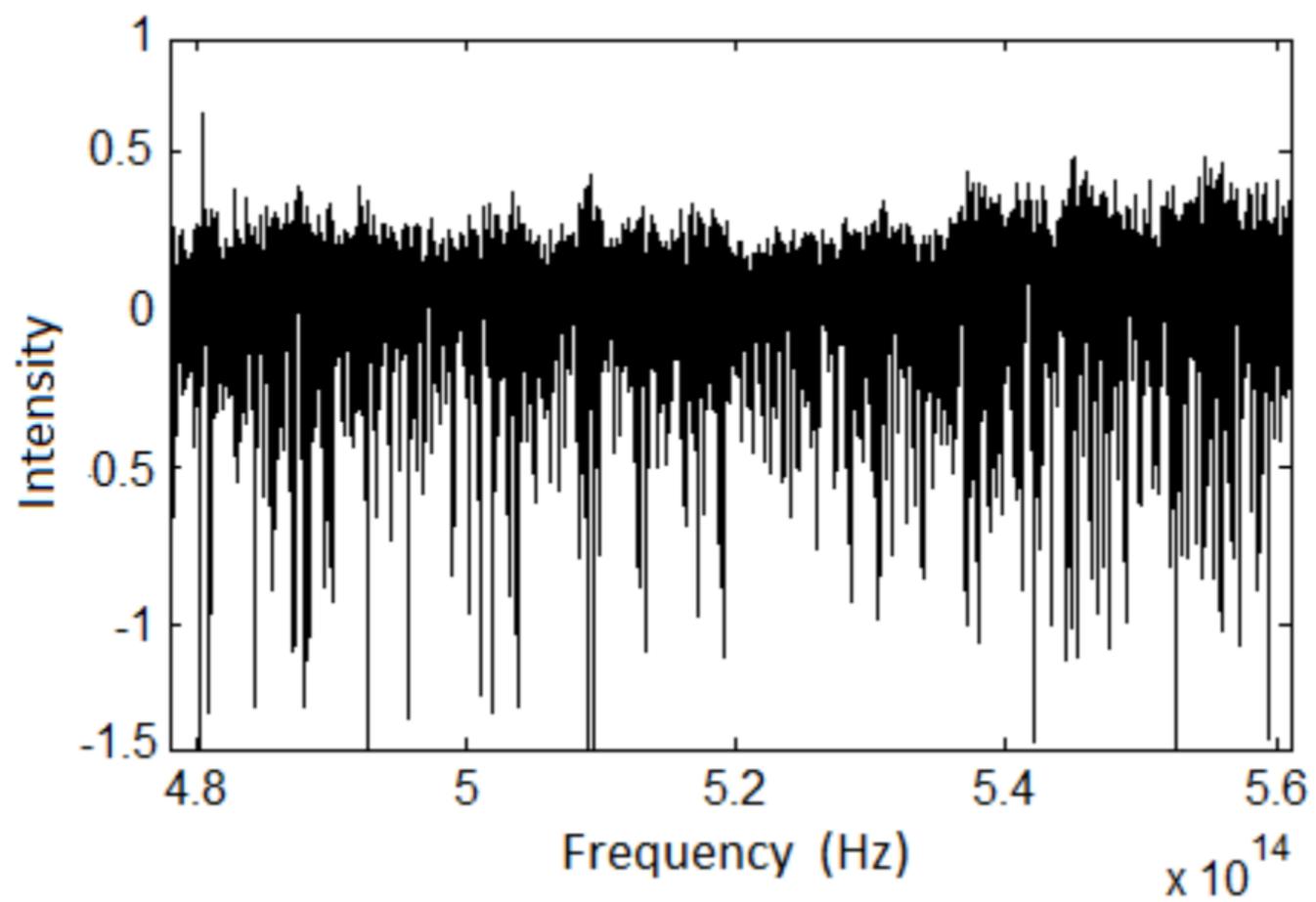 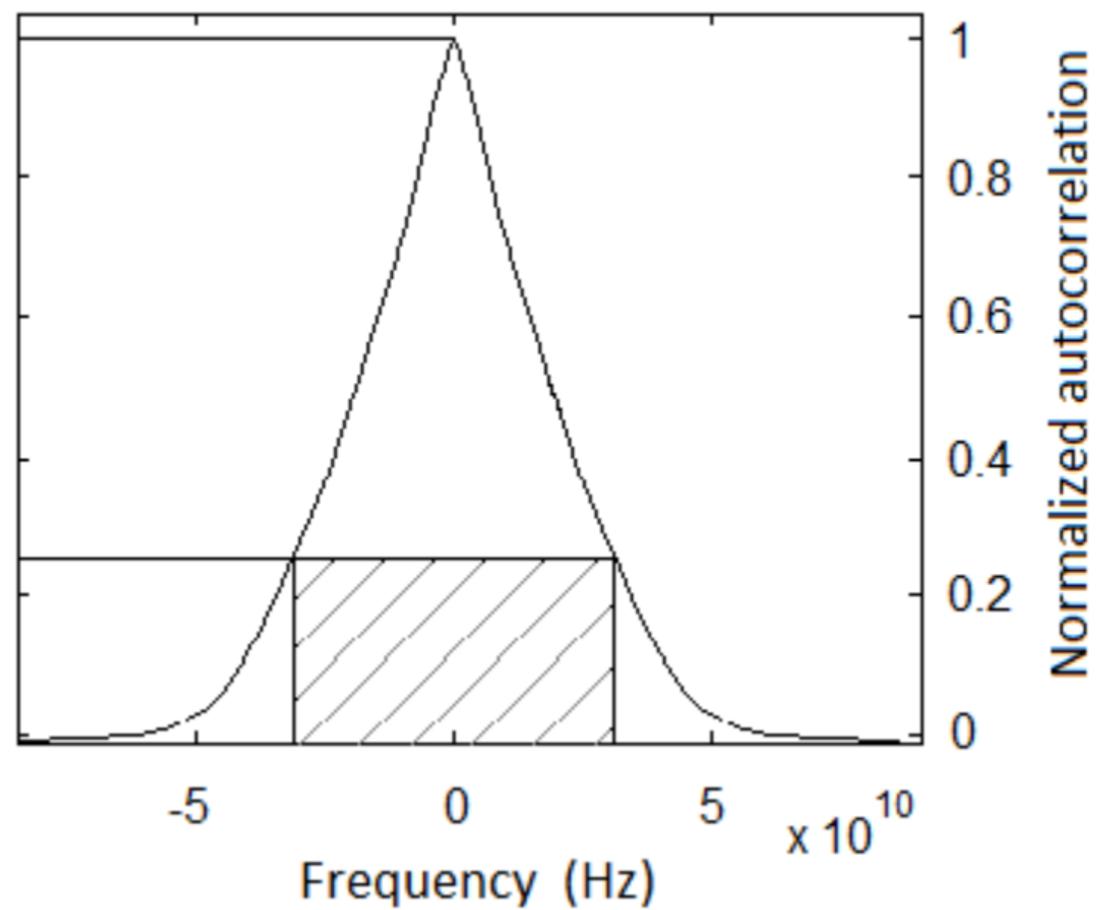

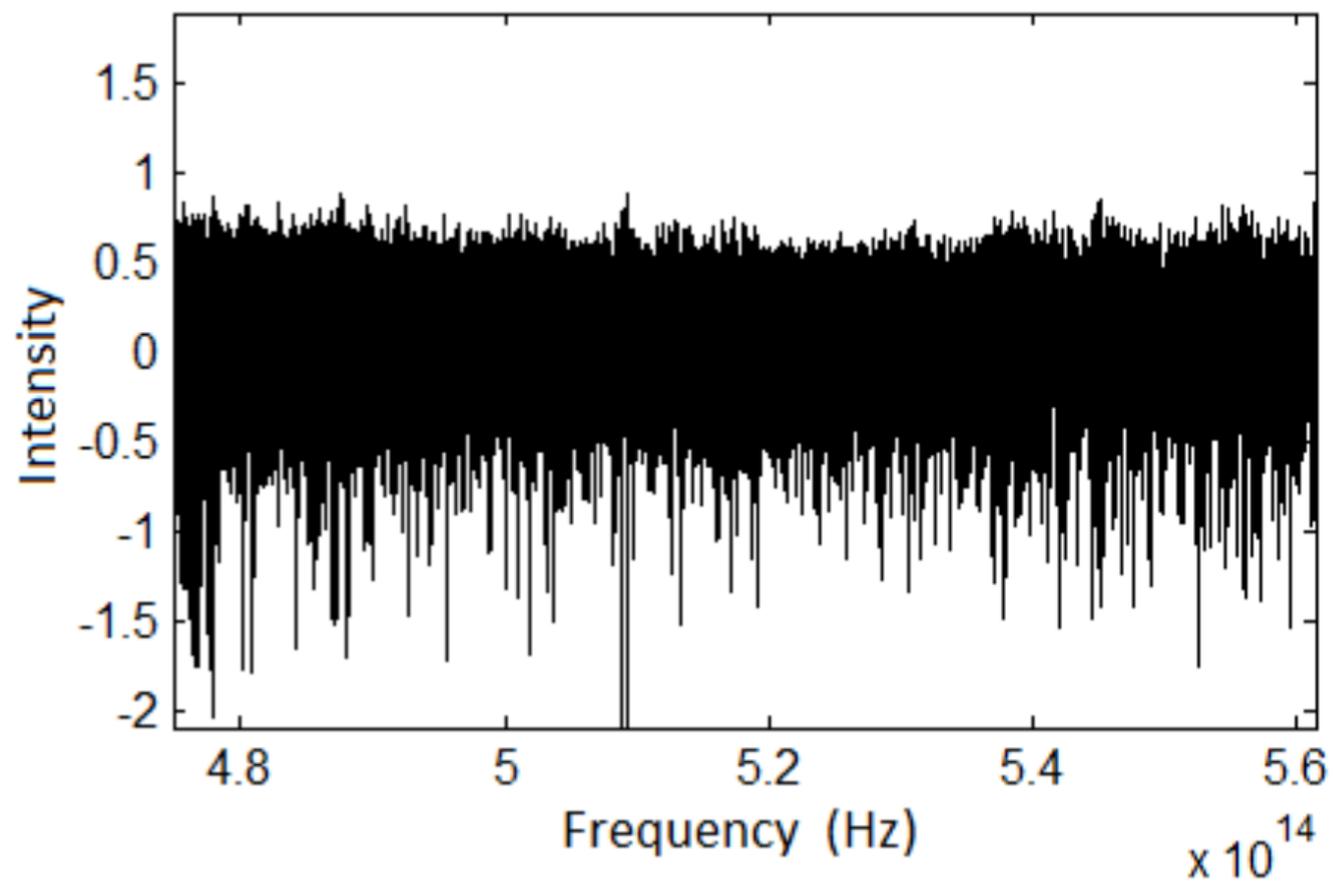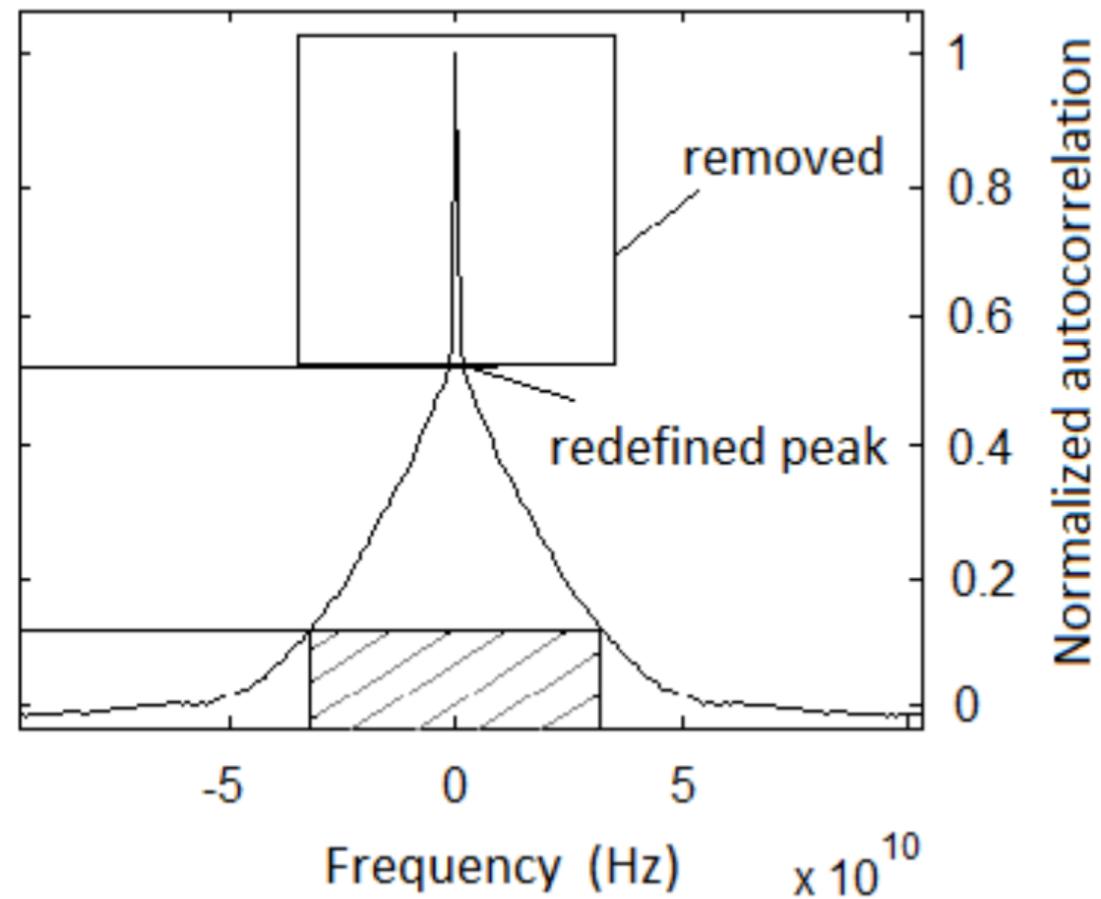

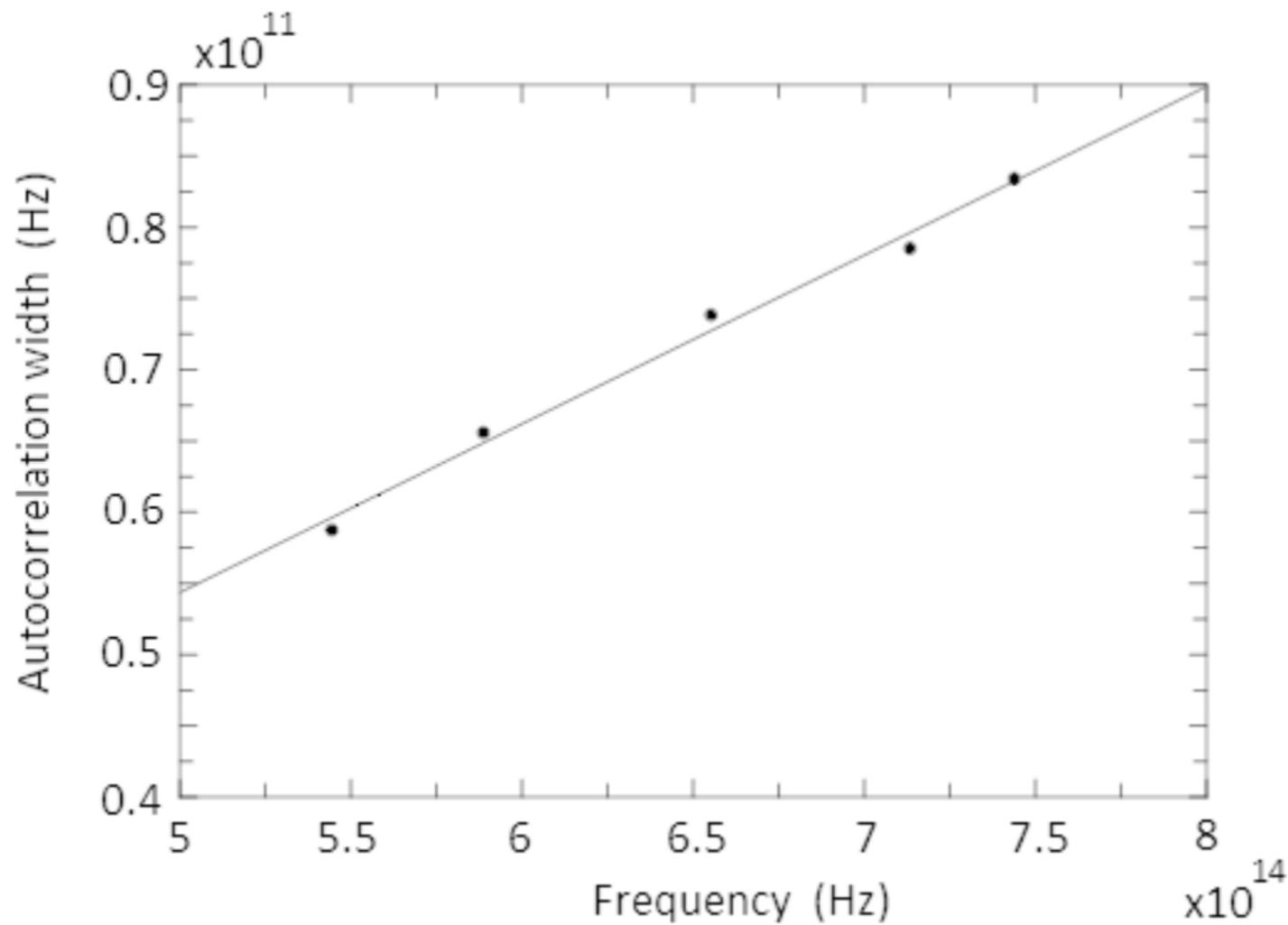

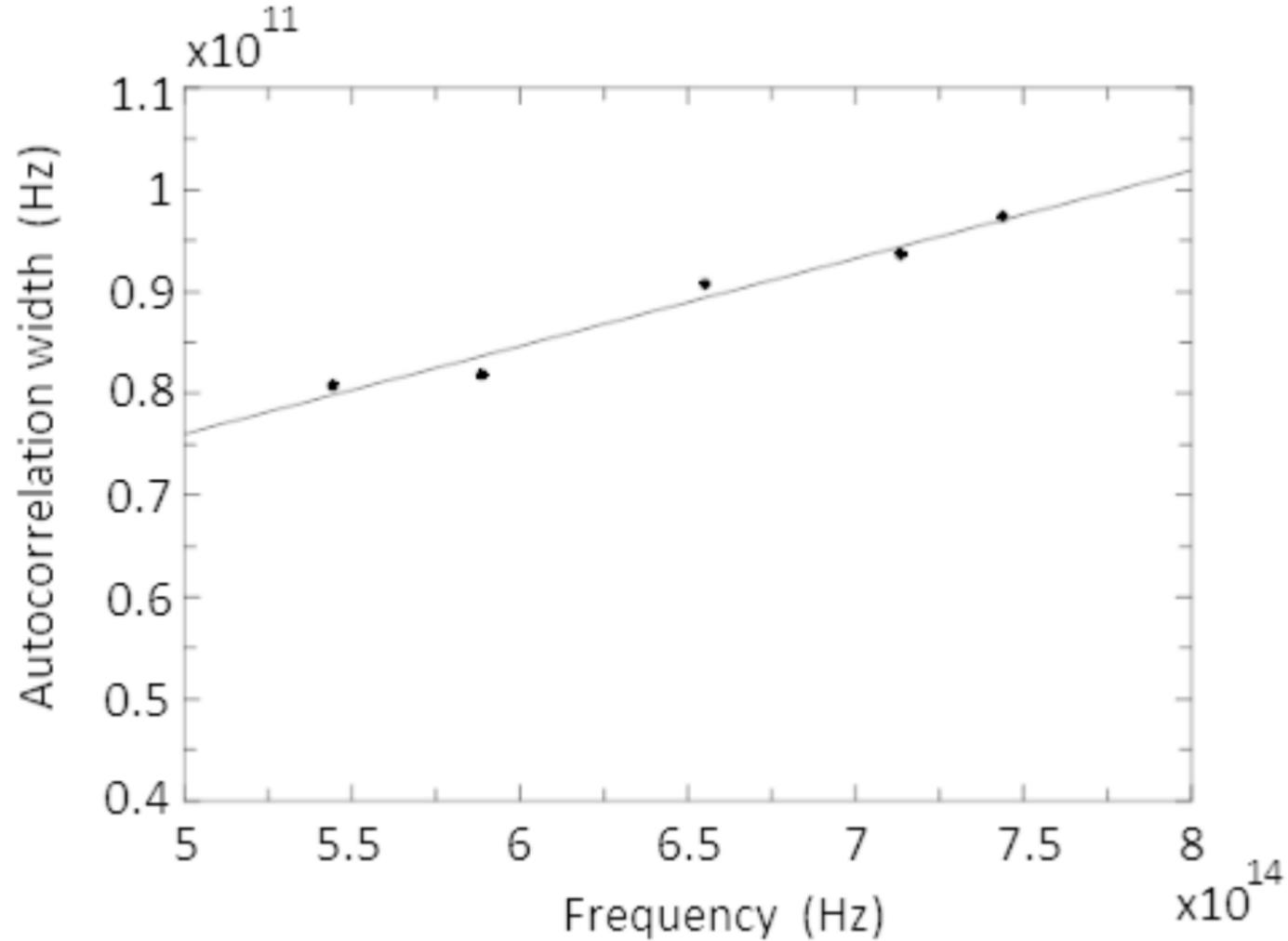

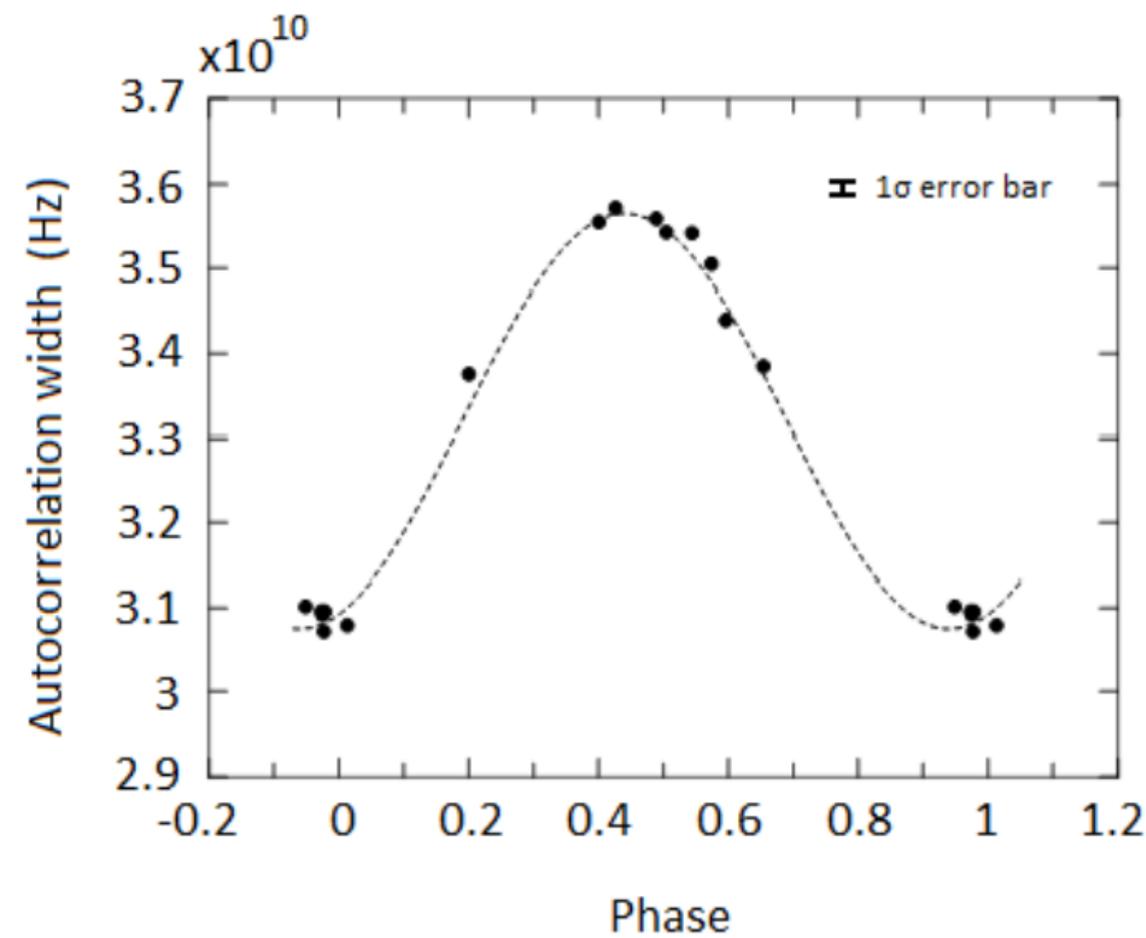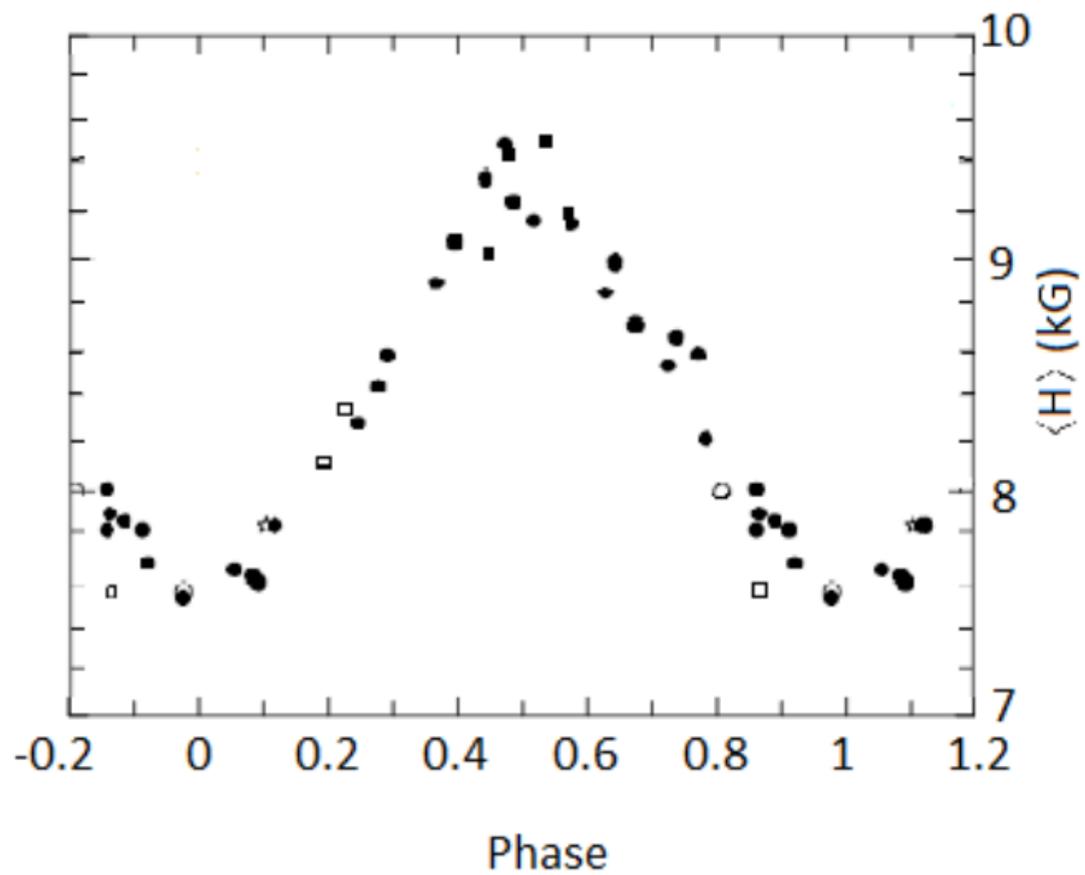

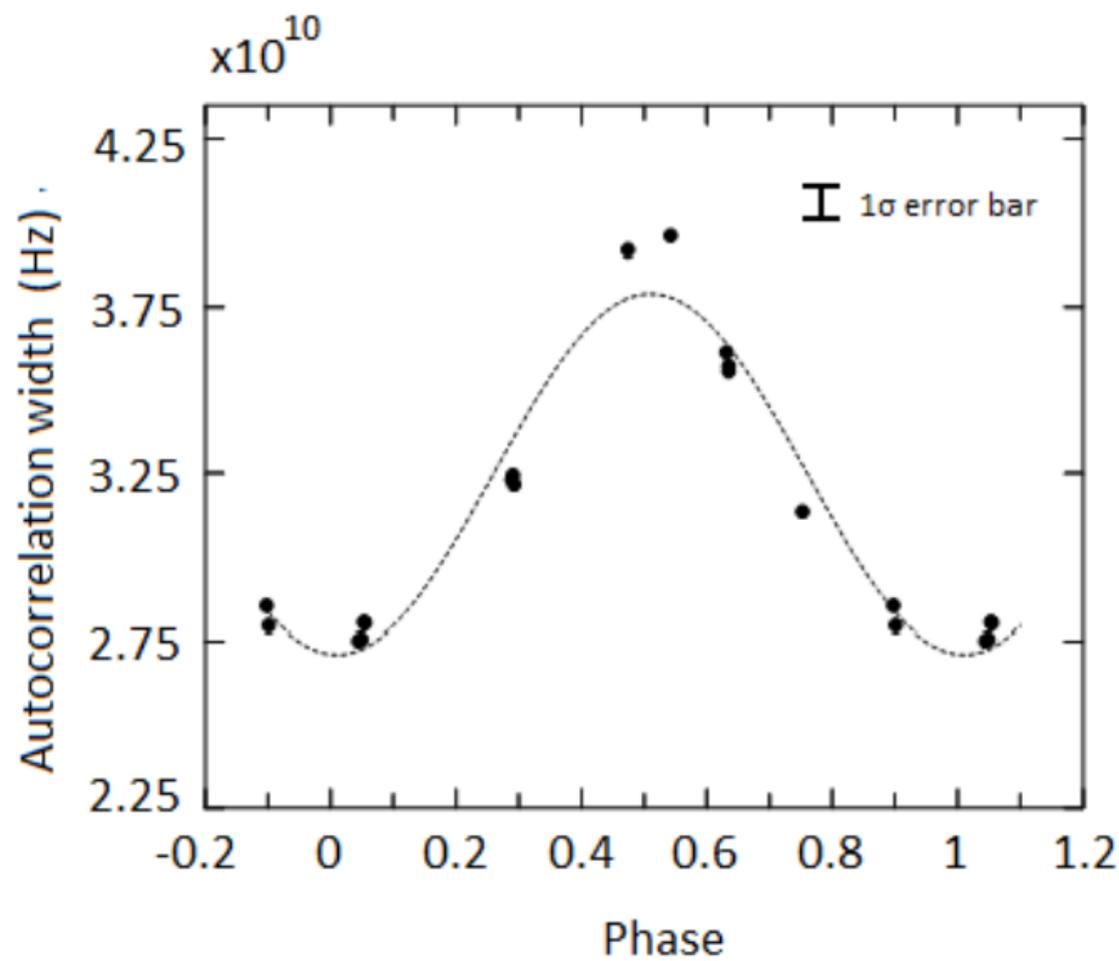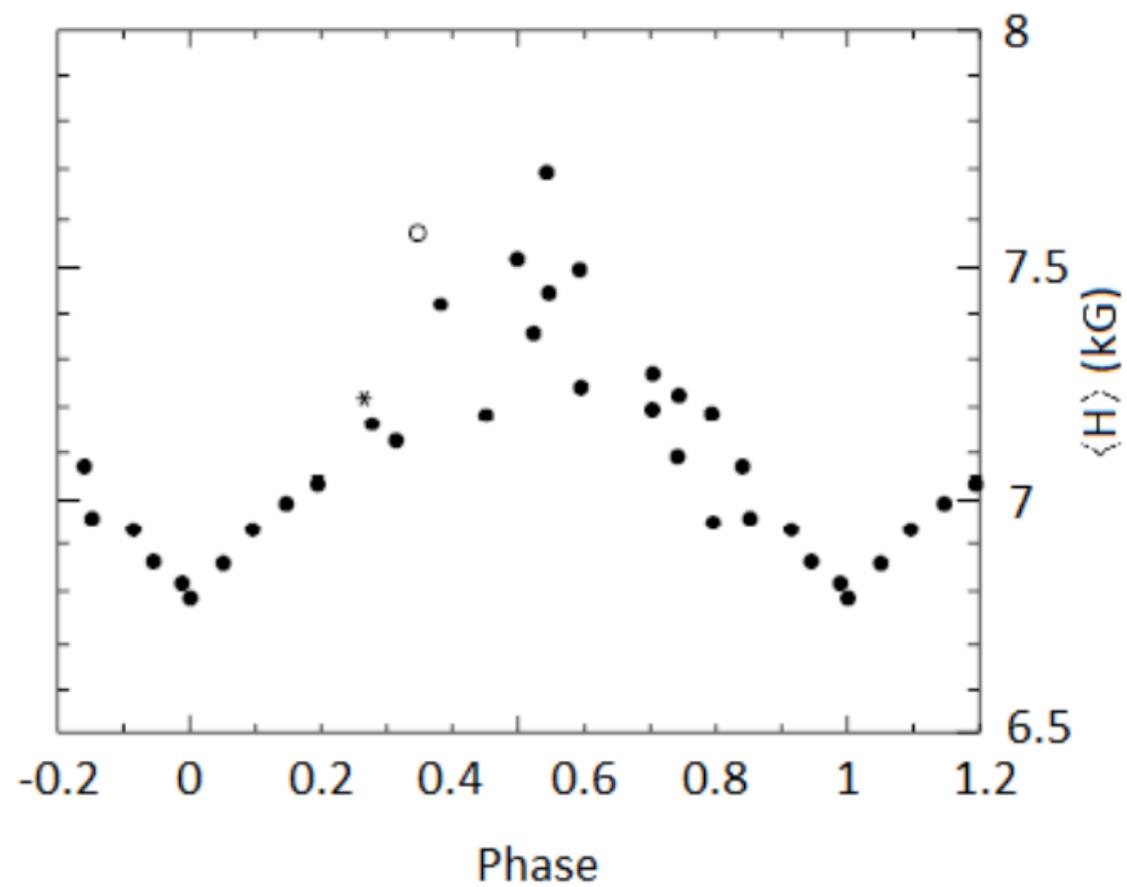

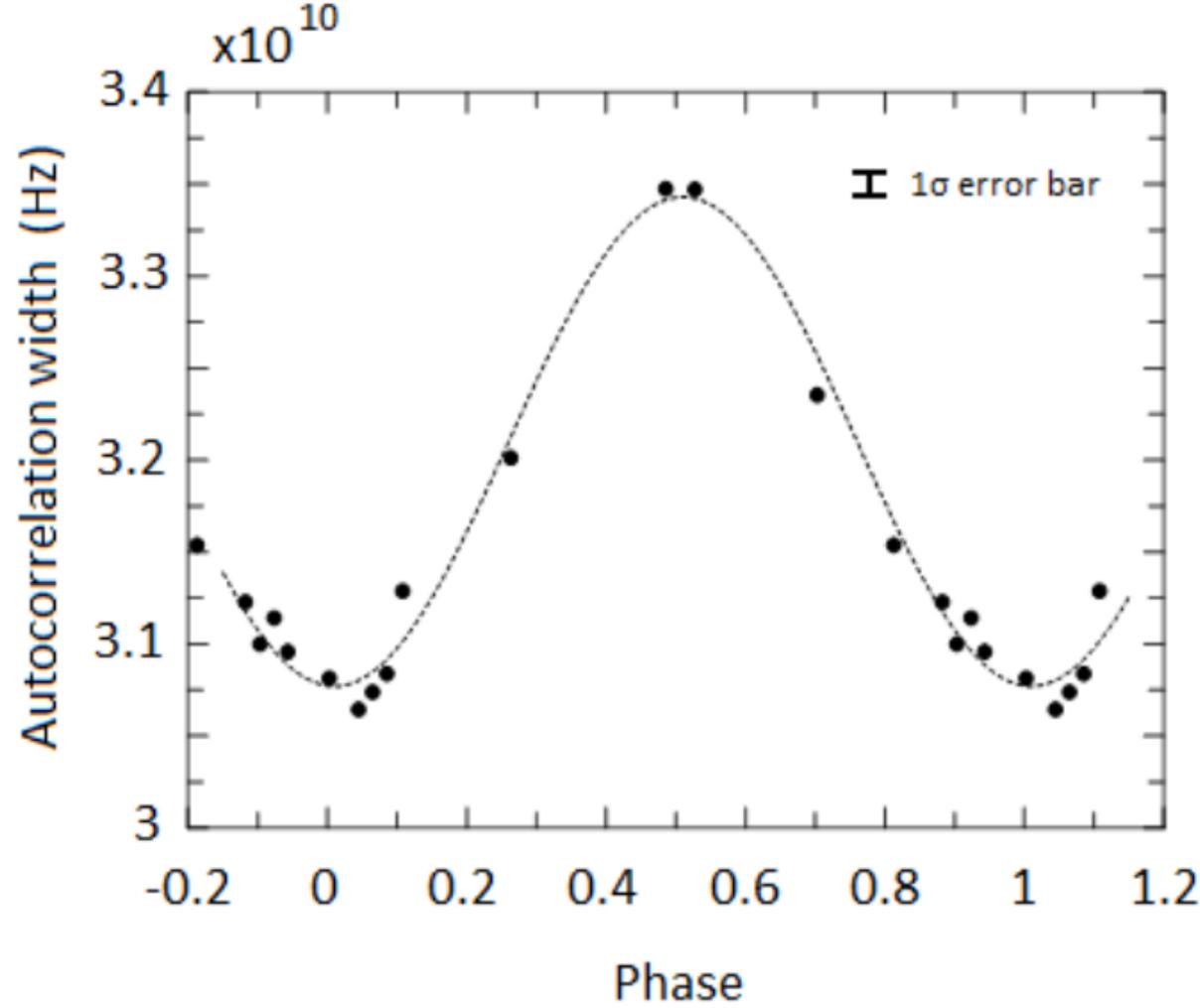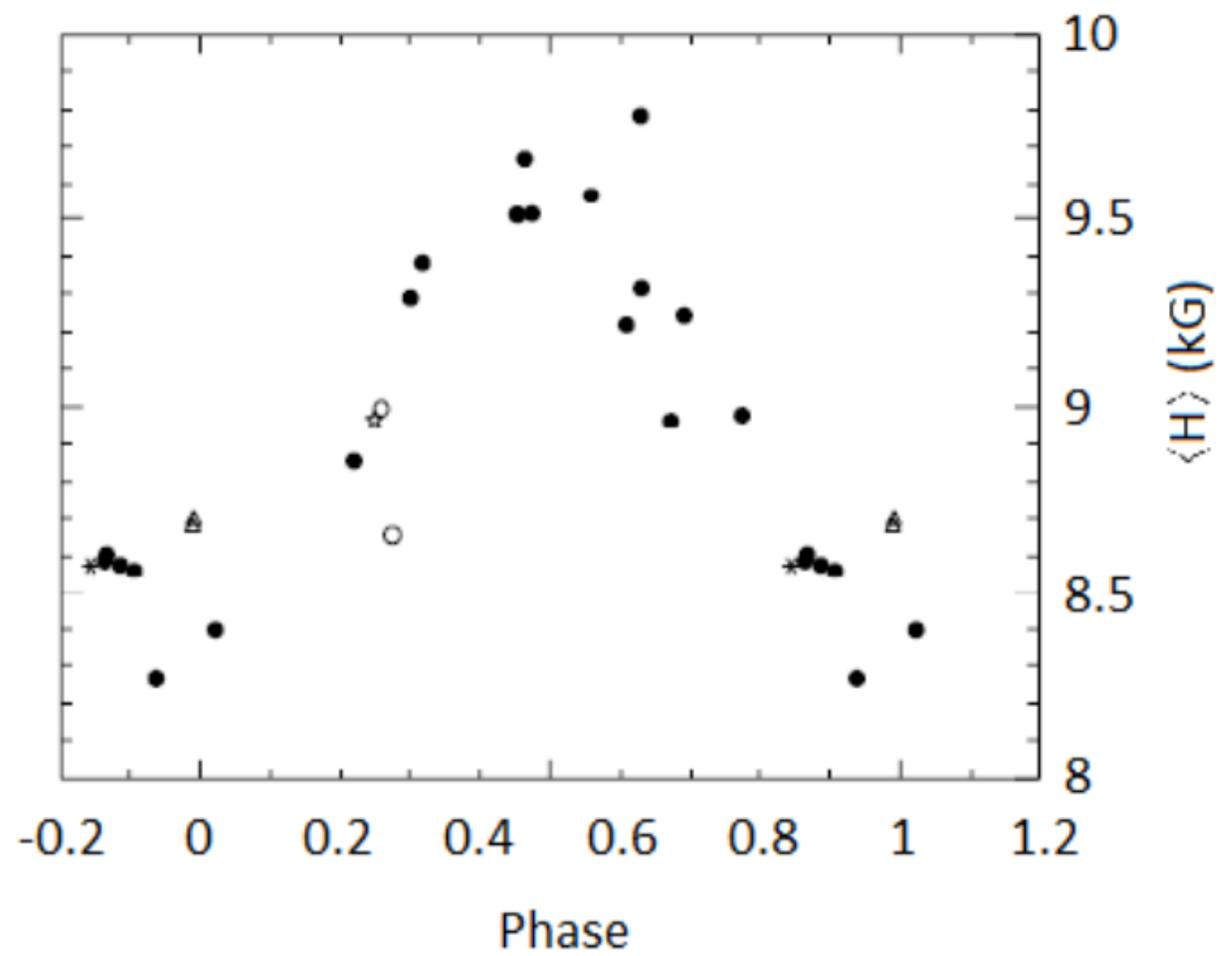

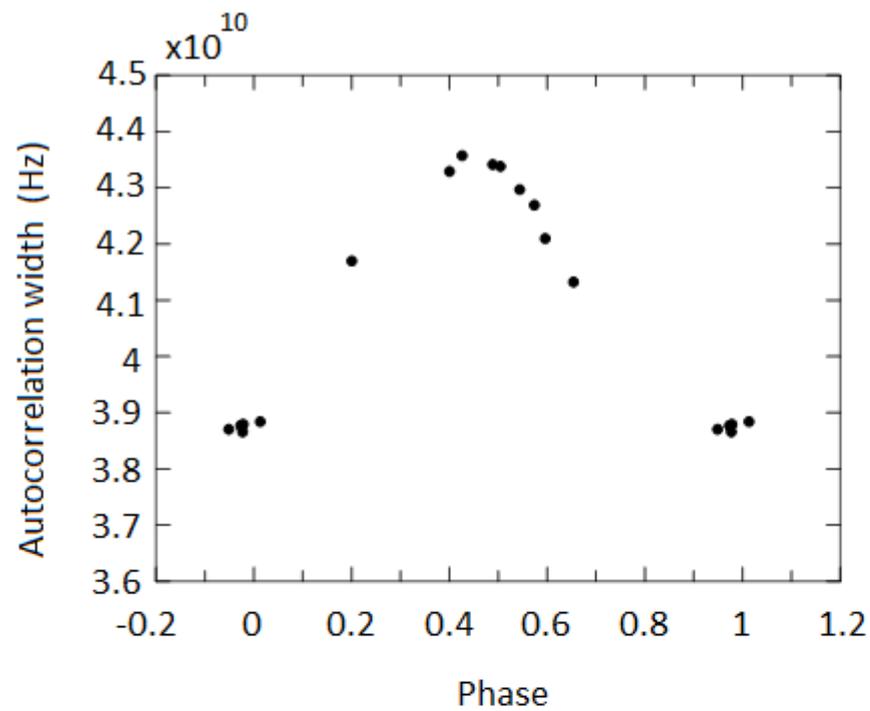 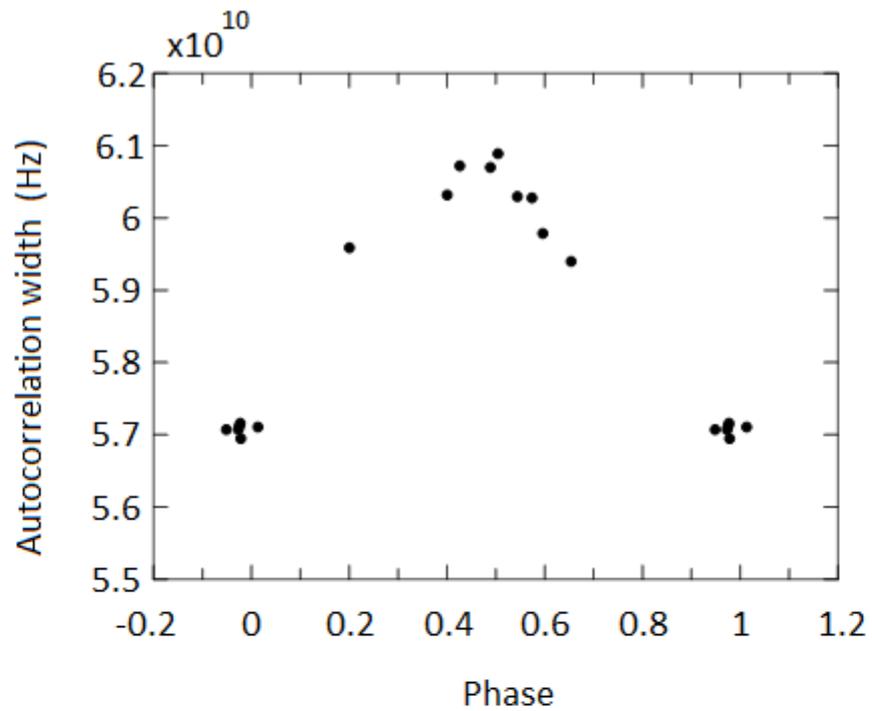 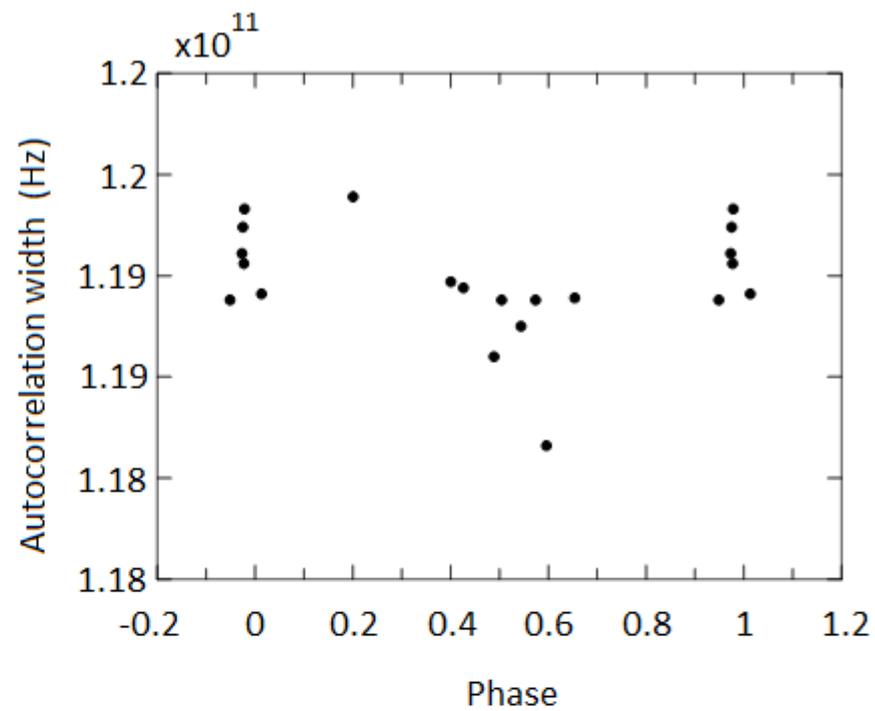
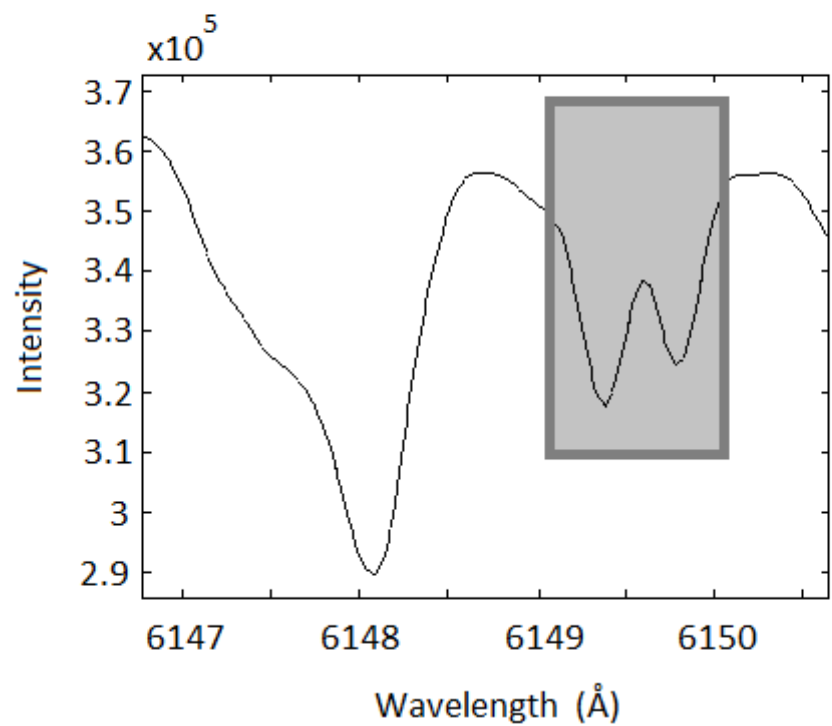 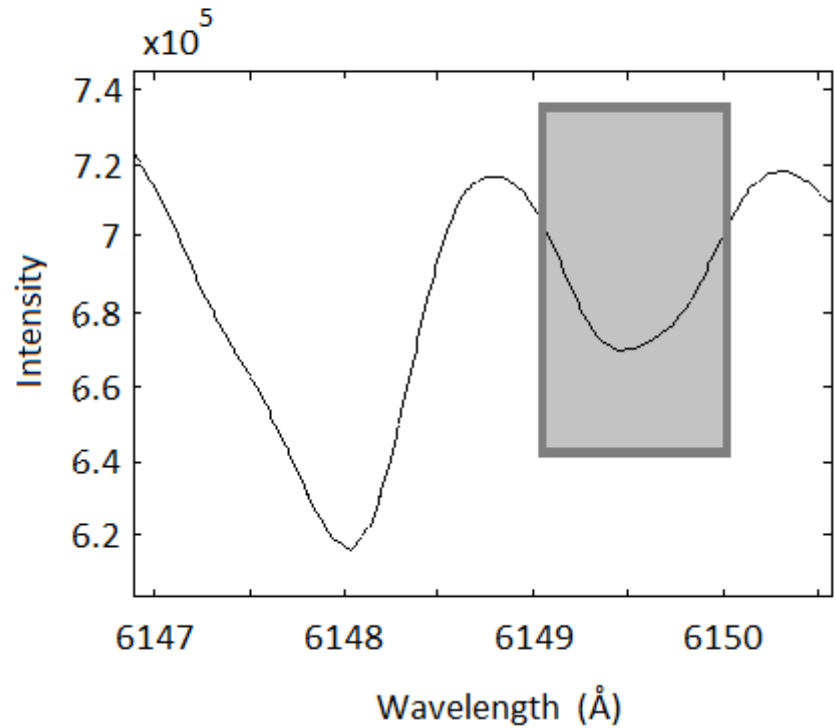 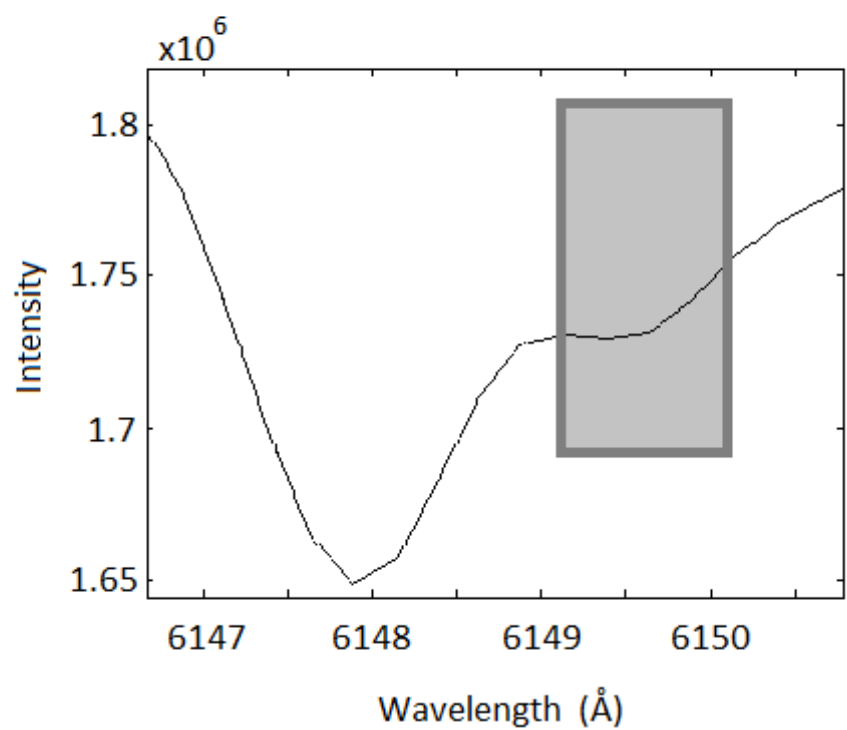

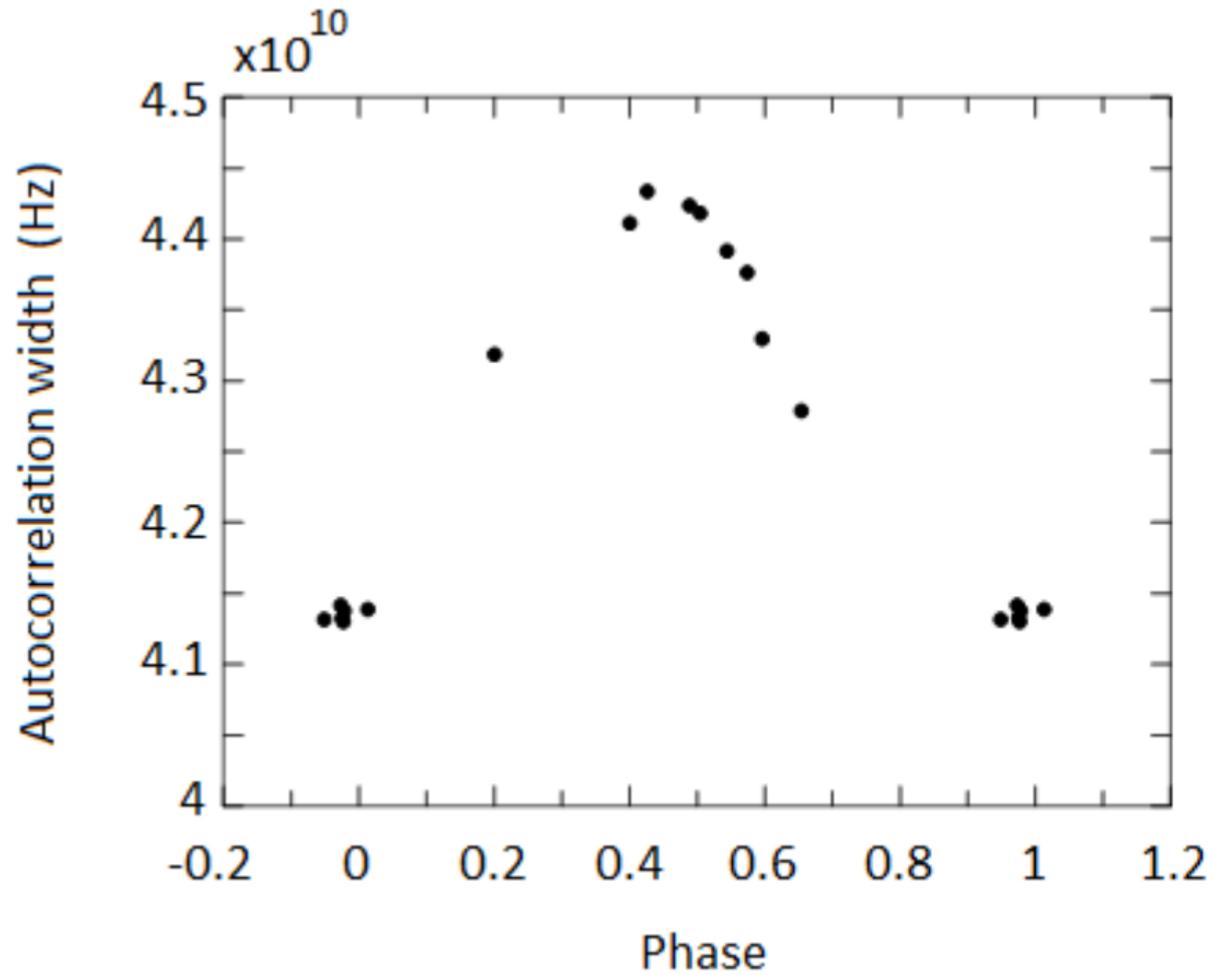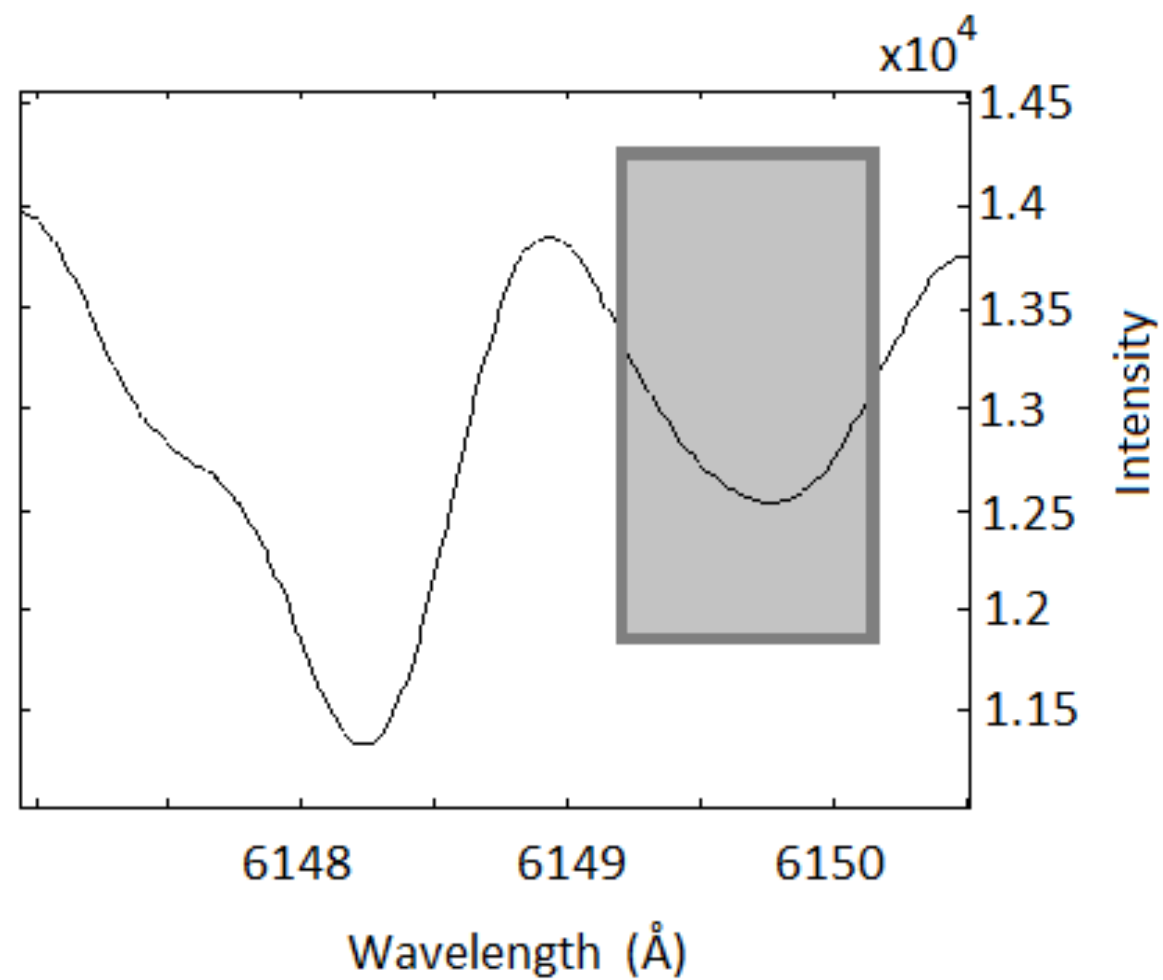

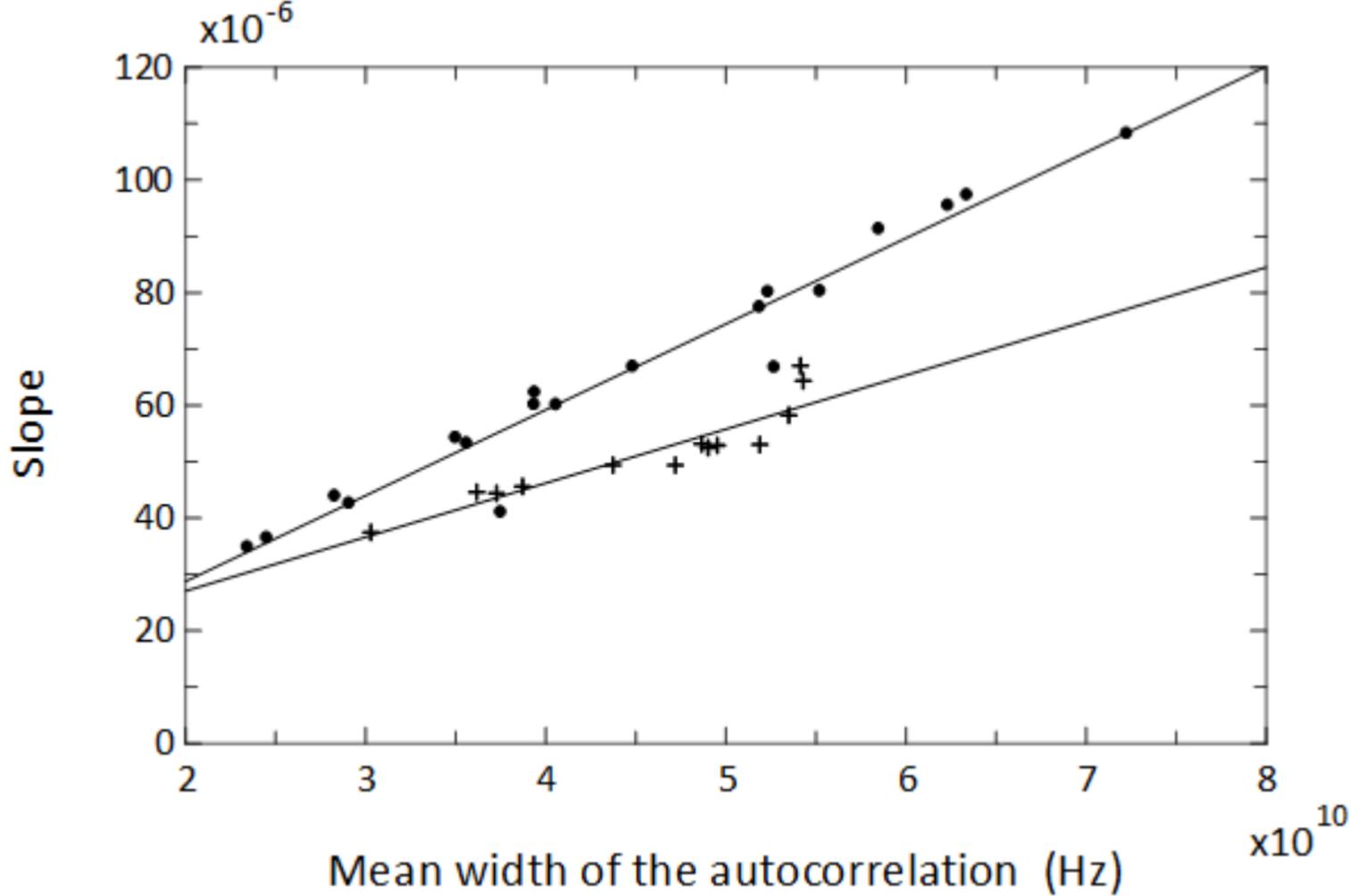

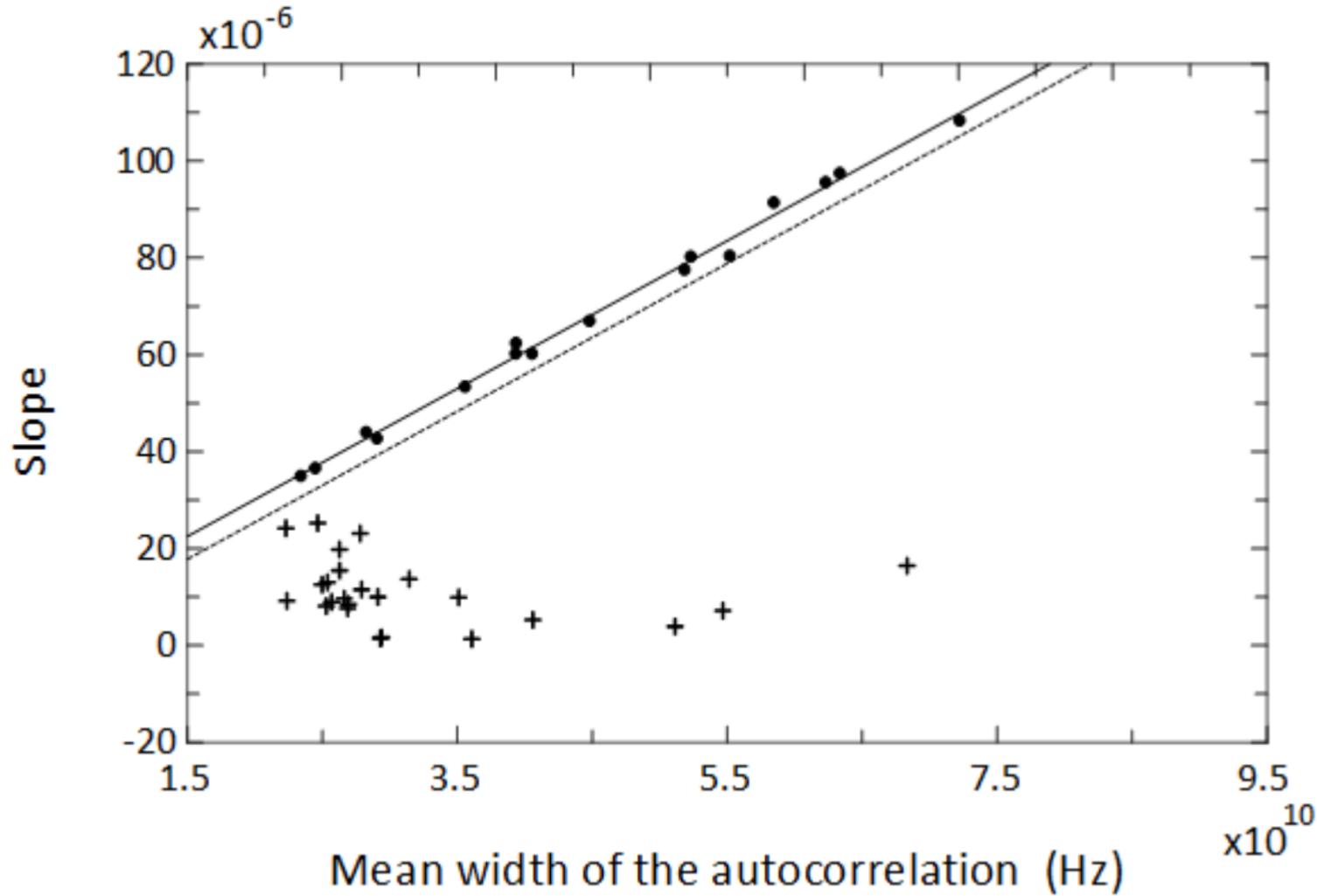

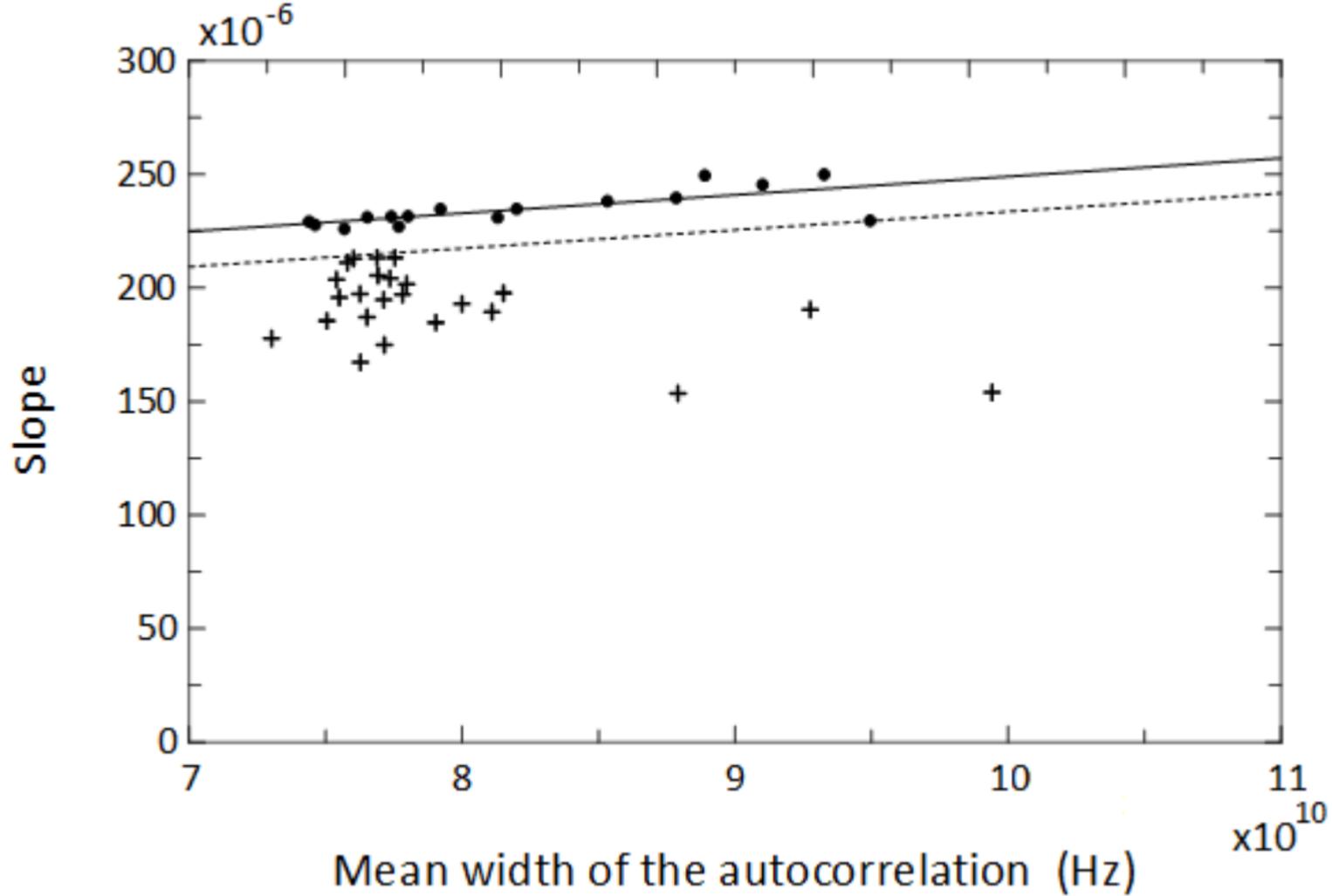

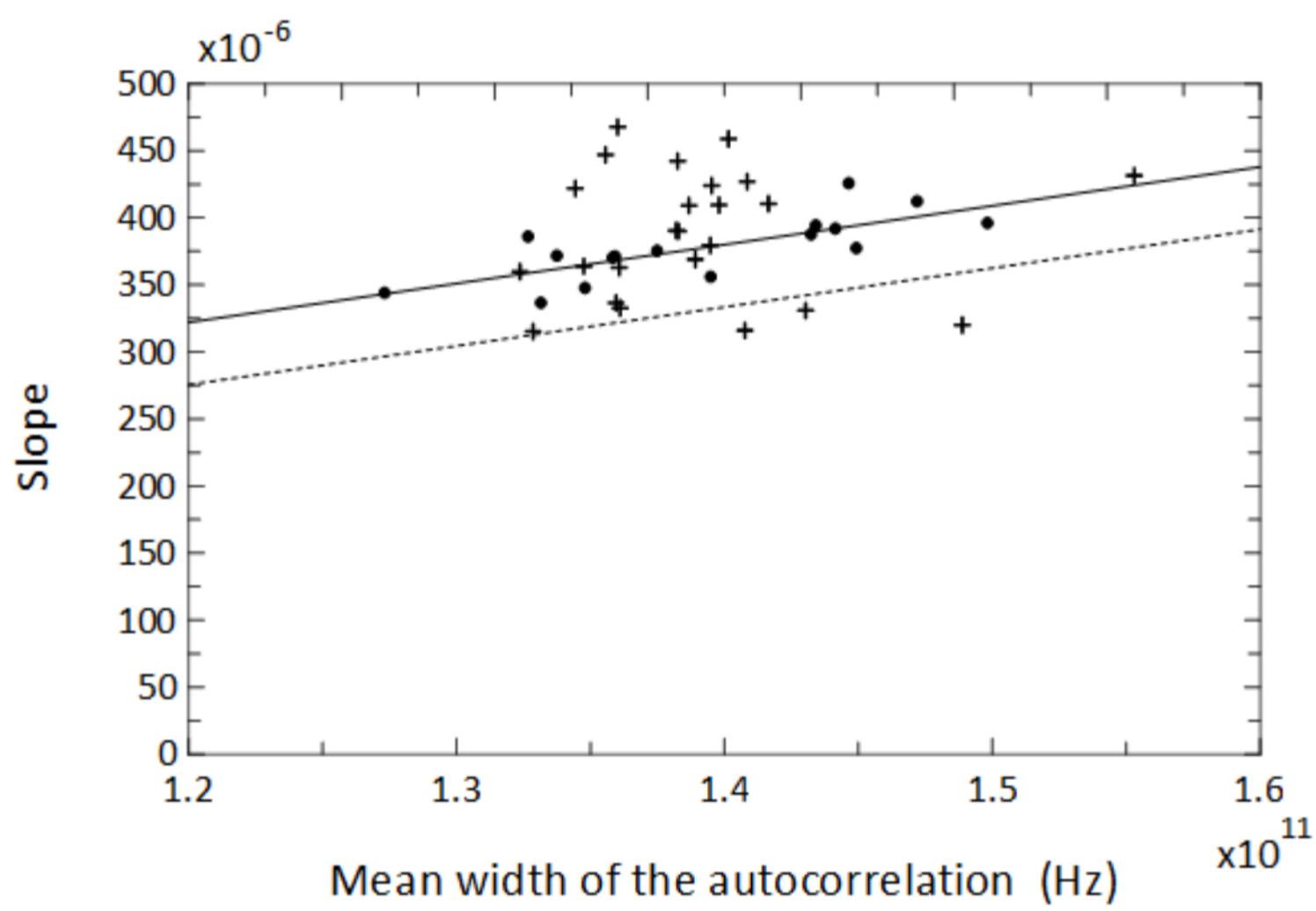

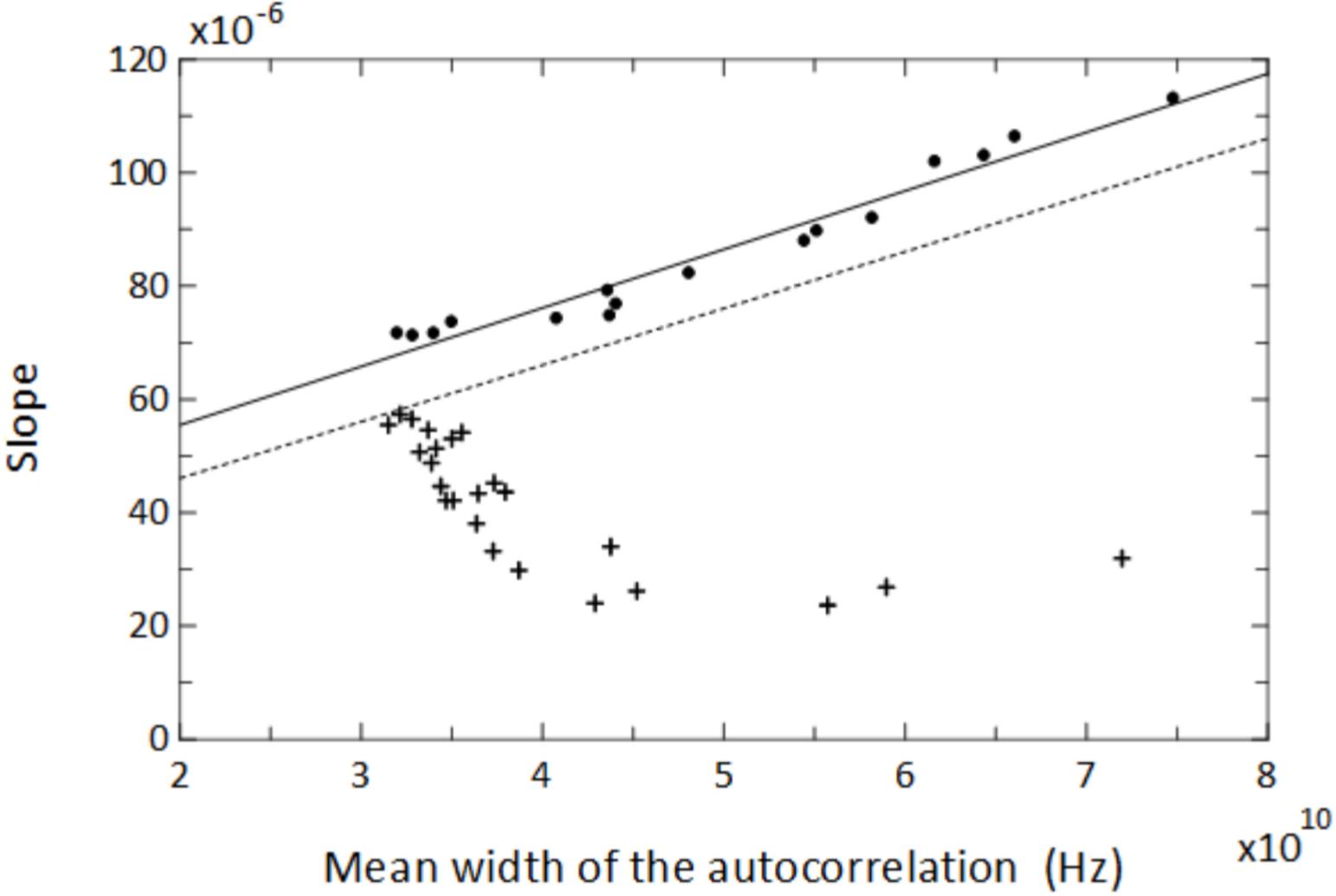

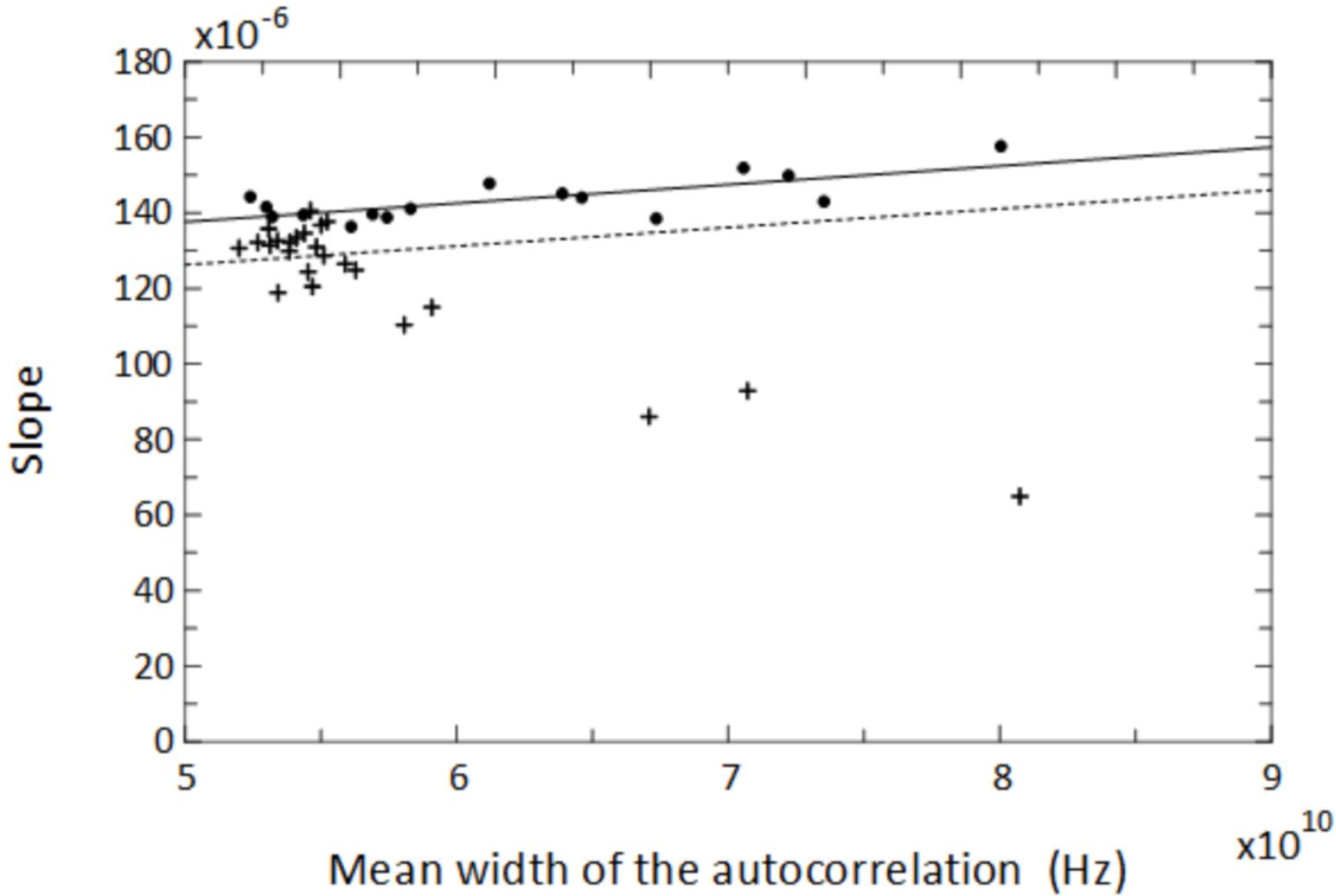